\documentclass[a4paper,aps,pre,floatfix,twocolumn,nofootinbib,showpacs,superscriptaddress]{revtex4-1}

\usepackage{times}
\usepackage{verbatim}
\usepackage{bbold}
\usepackage{bbm}
\usepackage[pdftex]{graphicx}
\setkeys{Gin}{width=0.9\columnwidth}
\usepackage{latexsym,amsmath,verbatim}
\usepackage{color}
\usepackage{hyperref}
\usepackage{lipsum}
\usepackage{rotating}
\usepackage{multirow}
\usepackage[english]{babel}
\usepackage{microtype}
\usepackage{comment}
\usepackage[normalem]{ulem}

\usepackage{color}
\hypersetup{colorlinks=true,
  linkcolor=blue,
  urlcolor=blue,
  citecolor=blue,
  pdfhighlight=/N}

\newcommand{\av}[1]{\langle{#1}\rangle}
\newcommand{\vphit}{\varphi_\mathrm{th}}

\begin{document}

\title{Scale-free behavioral cascades and effective leadership in
  schooling fish}

\author{Julia M\'ugica}

\affiliation{Departament de F\'isica, Universitat Polit\`ecnica de
  Catalunya, Campus Nord B4, 08034 Barcelona, Spain}

\author{Jordi Torrents}

\affiliation{Departament de F\'isica, Universitat Polit\`ecnica de
  Catalunya, Campus Nord B4, 08034 Barcelona, Spain}

\affiliation{Departament de F\'{\i}sica de la Mat\`{e}ria Condensada, Universitat
  de Barcelona, Mart\'{\i} i Franqu\`es 1, 08028 Barcelona, Spain}

\author{Javier Crist\'{\i}n}

\affiliation{Istituto Sistemi Complessi, Consiglio Nazionale delle
  Ricerche, UOS Sapienza, 00185 Rome, Italy}

\affiliation{Dipartimento di Fisica, Universita' Sapienza, 00185 Rome,
  Italy}

\author{Andreu Puy}

\affiliation{Departament de F\'isica, Universitat Polit\`ecnica de
  Catalunya, Campus Nord B4, 08034 Barcelona, Spain}

\author{M. Carmen Miguel}

\affiliation{Departament de F\'{\i}sica de la Mat\`{e}ria Condensada, Universitat
  de Barcelona, Mart\'{\i} i Franqu\`es 1, 08028 Barcelona, Spain}

\affiliation{Universitat de Barcelona Institute of Complex Systems (UBICS),
  Universitat de Barcelona, Barcelona, Spain}

\author{Romualdo Pastor-Satorras}
\affiliation{Departament de F\'isica, Universitat Polit\`ecnica de
  Catalunya, Campus Nord B4, 08034 Barcelona, Spain}

\begin{abstract}
  Behavioral contagion and the presence of behavioral cascades are
  natural features in groups of animals showing collective motion, such
  as schooling fish or grazing herbivores. Here we study empirical
  behavioral cascades observed in fish schools defined as avalanches of
  consecutive large changes in the heading direction of the trajectory
  of fish.  In terms of a minimum turning angle introduced to define a
  large change, avalanches are characterized by distributions of size
  and duration showing scale-free signatures, reminiscent of
  self-organized critical behavior. We observe that avalanches are
  generally triggered by a small number of fish, which act as effective
  leaders that induce large rearrangements of the group's
  trajectory. This observation motivates the proposal of a simple model,
  based in the classical Vicsek model of collective motion, in which a
  given individual acts as a leader subject to random heading
  reorientations. The model reproduces qualitatively the empirical
  avalanche behavior observed in real schools, and hints towards a
  connection between effective leadership, long range interactions and
  avalanche behavior in collective movement.
\end{abstract}

\maketitle

\section{Introduction}

Collective motion is an ubiquitous phenomenon in nature, observed in a
wide variety of different living systems and on an even wider range of
scales, from mammal herds and fish schools, to bacteria colonies and
cellular migrations~\cite{Vicsek2012,sumpter2010,Ramaswamy2010}. The
study of collective motion allows scientists to infer the intricate
interaction mechanisms governing the diversity of behaviors found in
natural grouping
species~\cite{Vicsek2012,Rosenthal2015,Chen2016,Calovi2018}. Identifying
the most relevant traits will prove essential if we ever want to take
advantage of nature wisdom for engineering applications such as in swarm
robotics~\cite{Brambilla2013} or in driver-less cars. Social animals
group and travel together to gain several benefits, from better foraging
and more efficient offspring training, to improved navigational accuracy
and reduced risk of predation~\cite{Krause2002}. Examples illustrating
the emergence of ordered collective motion in social animal groups can
take the spectacular form of wildebeest herds crossing deserts in
Africa, or huge fish schools running away coordinately from
predators~\cite{Procaccini2011}. From a more mundane perspective, the
seemingly simple movement of a sheep herd crossing a road also arises as
a result of the collective, coordinated motion of individual
sheep~\cite{Ginelli2015}.

A common view of collective motion, implemented in most numerical
models, is that coherent spatio-temporal patterns emerge spontaneously
from decentralized interactions among identical self-propelled group
members~\cite{Vicsek2012}. The kind of coordination required to produce
such impressive patterns, however,
requires an efficient transfer of information among the group
components. In this regard, leadership is sometimes brought about to
rationalize the cooperative movements of animal groups by single
individuals that appear to have a strong influence on the flock
behavior~\cite{Couzin2005}.  The effects of leadership have been
considered in several contexts, including crowd
behavior~\cite{Aube2004}, hierarchical
leadership~\cite{doi:10.1137/060673254}, linear response theory in
flocking systems~\cite{Pearce2016,Kyriakopoulos_2016} or the emergence
of complex patterns of cooperation and
conflict~\cite{Smith2016}. Leadership can arise as a natural instinct in
some animals, which form a permanent hierarchical structure, but in
other cases it can exhibit a switching dynamics that can even depend on
context~\cite{Nagy2010,Flack2012,Nagy2013,Chen2016}.  In this sense,
effective leadership can come from individuals having useful information
about their environment, such as the position of food or predators, not
visible to the rest of the flock~\cite{Couzin2011,Ward08022011}.
Another important aspect of collective animal motion is the existence of
spontaneous individual-level behavioral variations, which may be
transmitted to the group as if those particular individuals were
effective group leaders.  As a result of abrupt changes of dynamic
behavior at the level of one or a few individuals, animal groups can
exhibit intermittent collective rearrangements, or can even undergo
state transitions at the macroscopic level.  Collective behavioral
oscillations or waves in groups have been reported in golden
shiners~\cite{Rosenthal2015}, which were related to an underlying or
\textit{hidden} communication network~\cite{Rosenthal2015}.  On the
other hand, sheep herds have been shown to pass from slow group
dispersive motion while grazing, to rapid aggregation induced by sudden,
individual changes of speed~\cite{Ginelli2015} in the absence of nearby
threatening sources.  Most interestingly, these experimental studies
emphasize that animal rearrangements can either spread extensively
within the group or extinguish rapidly, leading to an avalanche-like
type of response with a broad-tailed distribution of avalanche
magnitudes~\cite{Rosenthal2015,Ginelli2015}. This sort of avalanche
behavior is well known in the physics literature~\cite{Fisher1998},
where it has been discussed in magnetic
materials~\cite{PhysRevB.58.6353},
superconductors~\cite{RevModPhys.76.471}, plastic deformation of
crystalline materials~\cite{Miguel2001}, fracture
phenomena~\cite{PhysRevE.59.5049}, or
earthquakes~\cite{RevModPhys.84.839}.

In this paper we examine the interplay between effective leadership and
behavioral cascades (avalanching behavior) by means of an empirical
analysis of the movement of black neon tetra fish \emph{Hyphessobrycon
  herbertaxelrodi}, and through the theoretical analysis of a variation
of the classical Vicsek model~\cite{vicsek1995} that includes an
explicit leader. In our empirical analysis, we define avalanches in
terms of changes in the fish heading above a given turning angle
threshold, which lead to a sudden reorientation of the global trajectory
of the school. We observe that the distributions of size and duration of
the measured avalanches show scale-free signatures in analogy with
self-organized critical processes~\cite{pruessner2012} that can be
described in terms of a set of characteristics scaling exponents. We
explore the possible presence of leadership by considering the
statistics of avalanche initiators, observing that some fish have an
anomalous large probability of starting an avalanche, acting thus as
effective leaders promoting substantial school rearrangements. In order
to check the general effects of leadership in avalanche behavior, we
consider a Vicsek-like model in which a global leader, which exerts a
long range influence over the group members, alternates a directed
motion, unaffected by other individuals, with sudden variations of its
direction of motion, in the spirit of run-and-tumble
locomotion~\cite{mendez14}. Akin to our experimental observations, the
model exhibits intermittent scale-free avalanche-like behavior, not
present in the original model. Our results confirm the presence of
scale-free signatures in behavioral cascades in collective
motion~\cite{Rosenthal2015} and highlight the role of effective
leadership and long range interactions in the emergence of this sort of
collective behavior.

\section{Results}

\subsection{Empirical analysis of schooling fish}

We have analyzed the avalanche behavior in black neon tetra
(\emph{Hyphessobrycon herbertaxelrodi}), a small freshwater fish (adult
mean body size of $2.5$cm) that have a strong tendency to form compact
and highly polarized schools~\cite{gimeno2016}. Experiments, performed
by the group of F. S. Beltr\'an and V. Quera at the Institute of
Neurosciences, University of Barcelona (Spain), consisted in groups of
$40$ individuals, freely swimming in an experimental rectangular tank of
dimensions $100 \times 93$cm and $5$cm of depth. Videos of the fish
movement were recorded at $20$ frames per second with a resolution of
$1072 \times 1004$ pixels.  Three independent recordings, each of length
$T=12000$ frames were performed. The path of individual fish was
digitized using a custom-made tracking software. The paths obtained were
later visually inspected to correct for a few anomalously large jumps,
due to the switching of fish identities in the tracking process.  The
trajectories obtained were finally smoothed by applying a Savitzky-Golay
filter of window length $7$ and polynomial order
$3$~\cite{doi:10.1063/1.4822961}. The final result were three digital
data series, labeled A, B and C, that we analyzed numerically.

\subsubsection{Avalanche analysis}

Supplementary Video SV~1 shows a rendering of a segment of the school
evolution in time for series A. The heading of each fish, marked by a
short arrow, is defined in terms of its instantaneous velocity
$\vec{v}_i(t)$.  Given the path of a fish as a function to time
$\vec{r}_i(t)$, $t=1, 2, \ldots, T$ (time measured in frames), we define
the velocity at time $t$ using a Richardson extrapolation of order $4$,
in terms of the expression~\cite{Fornberg1988}
\begin{equation}
  \vec{v}_i(t) = \frac{1}{12} \left [ \vec{r}_i(t-2)  - 8
    \vec{r}_i(t-1) +
    8 \vec{r}_i(t+1) -
    \vec{r}_i(t+2) \right].
    \label{eq:richardson}
\end{equation}

As we can see in Supplementary Video SV~1, fish tend to move in a
coherent and highly polarized fashion, swimming with a common and slowly
changing average velocity.  In this regime, heading variations are small
and rather smooth.  However, at some instants of time, we can recognize
swift rearrangements of the individuals' headings, that lead to a change
of the average orientation of the school, accompanied by an increase of
the average velocity and a decrease and a delayed increase of the global
order of the school (see the animated plot in SV~1).  We interpret these
sudden rearrangements of individual heading as triggeres of
\emph{avalanches} of activity. Avalanches are triggered at different
positions within the experimental tank and, in particular, they are also
initiated near the tank's walls, but they are not restricted to occur
always there.  In order to quantify them, we fist examine the angular
variations in the heading of individual fish, defined as the
\emph{turning angle} $\varphi_i(t)$ formed by the velocity vectors
$\vec{v}_i(t+1)$ and $\vec{v}_i(t)$, and computed as
\begin{equation}
  \varphi_i(t) = \left\lvert \arctan  \left \{  \frac{\lVert \vec{v}_i(t) \times
  \vec{v}_i(t+1) \rVert}{\lVert \vec{v}_i(t) \cdot \vec{v}_i(t+1) \rVert }
  \right\} \right \rvert,
  \label{eq:turning}
\end{equation}
where $\times$ stand for the vectorial product and $\lVert \cdot \rVert$
represents the vector modulus.  For symmetry reasons, the angles,
computed in the interval $[-\pi, \pi]$, are projected onto $[0, \pi]$.

\begin{figure}[t]
  \centering{\includegraphics[width=\columnwidth]{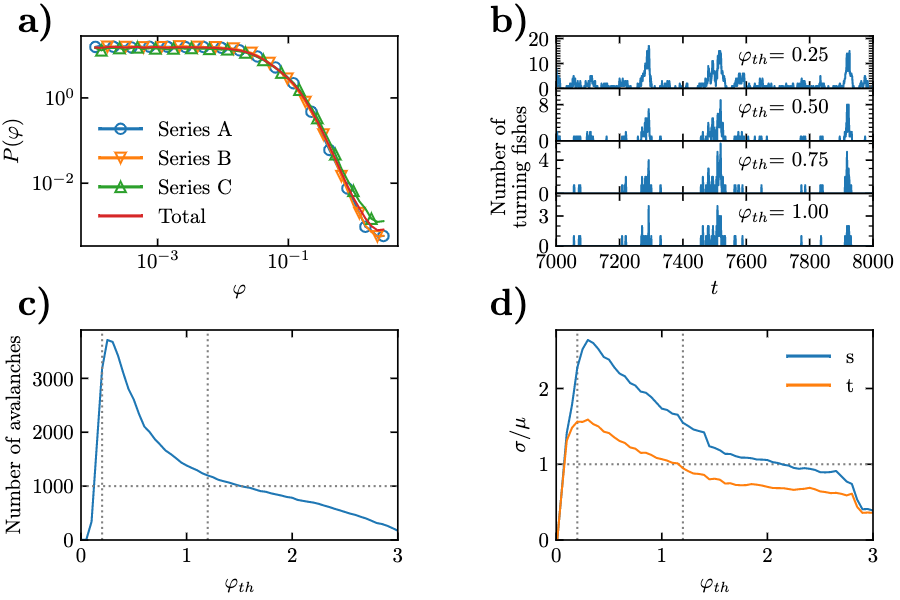}}

  \caption{(a) Probability density of turning angles $\varphi$ of
    individual fish. (b) Number of fish turning an angle larger than
    $\vphit$ in a sequence of $1000$ frames in data series A, for
    different values of the turning threshold. (c) Number of avalanches
    observed as a function of the turning threshold. (d) Relative
    fluctuations of the avalanche size ($s$) and duration ($t$)
    distributions as a function of the turning threshold. Vertical lines
    in panels (c) and (d) represent the turning threshold interval
    $[0.2, 1.2]$.}
  \label{fig:data_fish}
\end{figure}

In Fig.~\ref{fig:data_fish}(a) we plot the distribution of turning
angles $P(\varphi)$ for the different data series (hollow symbols) and
for the aggregation of all three of them (red line). Discarding some
extremely small values of $\varphi$, which can be attributed to
imprecision of the tracking algorithm when following an essentially
straight segment, the distributions show an extended plateau for small
turning angles, corresponding to stretches of time in which fish barely
change their heading and are thus compatible with movement along a
smoothly winding trajectory. Instead, for values larger than $0.01$
radians, the distribution starts to decrease sharply. These large turns
correspond to the sparsely distributed large rearrangements of direction
observed in SV~1.  In order to quantitatively identify avalanches, as is
customarily done in condensed matter physics~\cite{Laurson2006}, we
define a \emph{turning threshold} $\vphit$ that distinguishes small
turns $\varphi < \vphit$, associated to smooth trajectories, from large
turns $\varphi > \vphit$, associated to sudden rearrangements that
trigger an avalanche. In Fig.~\ref{fig:data_fish}(b) we plot, for a
given value of the threshold, the number of \emph{active} fish, defined
as those performing a turn larger than $\vphit$, as a function of time.
Here we can see the actual presence of turning avalanches, defined as
trains of consecutive frames in which more than one fish is active,
delimited by two frames (one at the beginning and another at the end of
the train) with no active fish.  These curves highlight the intermittent
and heterogeneous character of avalanches, which may be rather small or
can also reach relatively large sizes.

Our experimental data only allows the identification of a limited number
of avalanches. Indeed, in Fig.~\ref{fig:data_fish}(c) we plot the total
number of recorded avalanches as a function of the turning
threshold. From here we observe that the range of values of $\vphit$
that lead to at least $1000$ avalanches range approximately in the
interval $[0.20, 1.50]$. To study the statistics of avalanches, we
compute their duration $t$ and size $s$, defined as the number of
consecutive time steps (frames) with at least one active fish, and the
sum of the number of active fish at each time step of an avalanche,
respectively. Notice that, since a fish can be active in more than
  one step along the duration of an avalanche, the avalanche size $s$ is
  in general larger than the avalanche duration $t$, and can be larger
  than the total number of fish in the experiment.  A first broad statistical
characterization of avalanches is given by the relative size and
duration fluctuations, measured as the standard deviation $\sigma$
divided by the corresponding average value $\mu$.  In
Fig.~\ref{fig:data_fish}(d), we plot these relative fluctuations for
both $s$ and $t$, respectively. From this plot, we observe that relative
fluctuations are only larger than $1$ for threshold values within the
interval between $0.1$ and $1.2$. We therefore restrict our analysis to
the conservative threshold interval $[0.20, 1.20]$.

\begin{figure}[t]
  \centering{\includegraphics[width=\columnwidth]{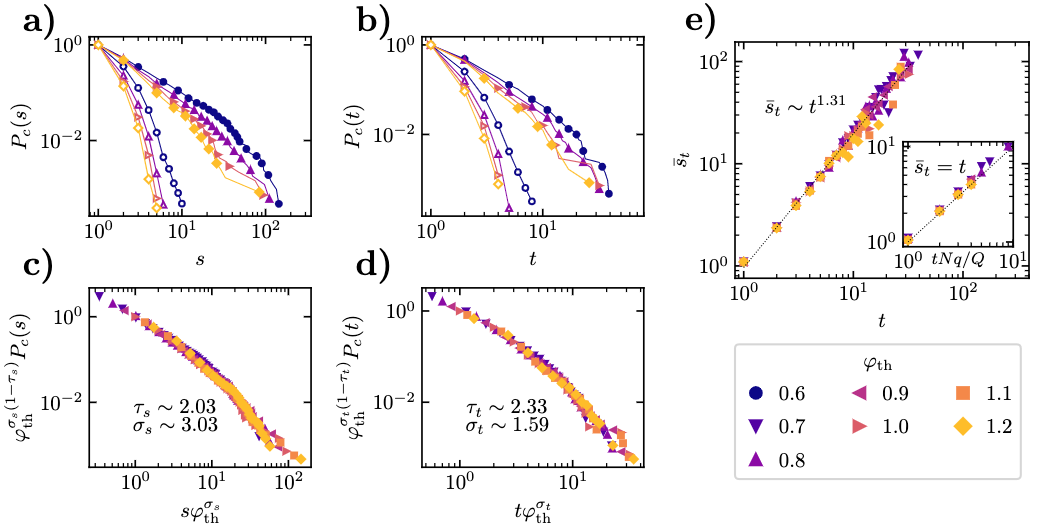}}

  \caption{\label{fig:distributions_fish} (a) Cumulative probability
    distribution of the avalanche sizes $P_c(s)$ for different values of
    the turning threshold $\vphit$. (b) Cumulative probability
    distribution of the avalanche durations $P_c(t)$ for different
    values of the turning threshold $\vphit$.  In (a) and (b), filled
    symbols represent the actual empirical distributions, while hollow
    symbols correspond to distributions obtained by randomizing the
    turning angles in the trajectory of each fish. (c) Check of the
    scaling of the cumulative size distribution with the turning
    threshold, as given by Eq.~\eqref{eq:collapse}. (d) Check of the
    scaling of the cumulated time distribution with the turning
    threshold, as given by Eq.~\eqref{eq:collapse}.  (e) Average size
    $\bar{s}_t$ of avalanches of fixed duration $t$ as a function of
    $t$. The main plot shows the empirical data. The inset presents the
    results from a randomization of the turning angles in each fish
    trajectory. In this case, we plot the average duration as a function
    of the theoretical prediction $N q t/Q$, see
    Eq.~(\ref{eq:mean_size_randomized}).}

\end{figure}

We consider the shape of the probability distributions of avalanche
sizes, $P(s)$, and durations, $P(t)$, focusing on the cumulative
distributions,
\begin{equation}
  P_c(s) = \sum_{s' = s}^\infty P(s') \qquad \mbox{and} \qquad P_c(t) =
  \sum_{t' = t}^\infty P(t').
  \label{eq:distrib}
\end{equation}
In Fig.~\ref{fig:distributions_fish}(a) we plot in full symbols the
cumulative size distribution obtaeeined for different values of $\vphit$.
From the double logarithmic scale in the plot, we can see that the size
distributions show long tails, compatible with a power-law behavior of
the form $P(s) \sim s^{-\tau_s}$ for small values of $s$. This power-law
behavior is due to the correlated nature of turns in the fish school, a
feature that can be corroborated from the comparison of these results
with the avalanche distributions obtained from trajectories
reconstructed by randomizing the sequence of turning angles of each
fish. In the latter case, one obtains a clear exponential decay, as
shown in hollow symbols in Figs.~\ref{fig:distributions_fish}(a)
and~(b); see Methods for an analytical derivation.

Upon closer scrutiny, we can also observe that, for sufficiently large
$\vphit$, the initial power-law behavior of the size distributions is
followed by a faster decay for $s$ larger than a characteristic size
$s_c$ that appears to be a decreasing function of the threshold
$\vphit$. Inspired by the observations in other avalanche
systems~\cite{Laurson2009} and in models of self-organized
criticality~\cite{pruessner2012}, we can assume that, for different
values of the threshold, the size distributions exhibit a scaling
behavior of the form
\begin{equation}
  P(s) = s^{-\tau_s} G_s\left( \frac{s}{s_c(\vphit)}\right )
  \label{eq:scaling}
\end{equation}
where the scaling function $G_s(z)$ is constant for small $z\ll 1$ and
decays rapidly to zero for $z \gg 1$. In analogy with avalanches in
condensed matter and critical phenomena~\cite{yeomans,cardy_1996} we
make the ansatz for the behavior of the size cut-off
$s_c(\vphit) \sim \vphit^{-\sigma_s}$, where $\sigma_s$ is a
characteristic exponent. We can estimate the values of the exponents by
noticing that Eq.~\eqref{eq:scaling} implies, for the cumulative
distribution,
$P_c(s) = s^{-\tau_s + 1} F_s \left( s \vphit^{\sigma_s} \right )$,
where $F_s(z)$ is another scaling function. The previous expression can
be rewritten as
\begin{equation}
  \vphit^{\sigma_s(1-\tau_s) }P_c(s) =  F'_s \left( s \vphit^{\sigma_s} \right ),
  \label{eq:collapse}
\end{equation}
where $F'_s(z) = z^{-\tau_s +1} F_s(z)$.  Eq.~\eqref{eq:collapse}
implies that, when plotting the rescaled distribution
$\vphit^{\sigma_s(1-\tau_s) } P_c(s)$ as a function of the rescaled size
$s \vphit^{\sigma_s}$, with the correct exponents $\tau_s$ and
$\sigma_s$, plots for different values of $\vphit$ should collapse onto
the same universal function $F'_s(z)$.  Using this idea, one can
estimate numerically the exponents $\tau_s$ and $\sigma_s$ as those that
provide the best collapse of the data rescaled using
Eq.~\eqref{eq:collapse} for the different values of $\vphit$, see
Methods.

Following this approach, using values of
$\vphit \in \{ 0.7, 0.8, 0.9, 1.0, 1.1, 1.2\}$, we estimate the
exponents $\tau_s \simeq 2.03$ and $\sigma_s \simeq 3.03$. In
Fig.~\ref{fig:distributions_fish}(c) we show the data collapse for
Eq.~\eqref{eq:collapse} obtained for the cumulated size distributions
using these values. Different intervals of the turning threshold provide
slightly different values of the exponents, from which we estimate the
average exponents quoted in Table~\ref{tab:exponents}. The same
procedure can be applied to the duration distribution, see
Fig.~\ref{fig:distributions_fish}(b), where now the cumulative duration
distribution $P_c(t)$ fulfills Eq.~\eqref{eq:collapse} with the
corresponding exponents $\tau_t$ and $\sigma_t$. In the same interval of
thresholds we find $\tau_t \simeq 2.33$ and $\sigma_t \simeq 1.59$, see
Fig.~\ref{fig:distributions_fish}(d), while the average exponents are
given in Table~\ref{tab:exponents}.  We can check the validity of these
results considering that, for small values of $s$ and $t$, the
distributions $P(s) \sim s^{-\tau_s}$ and $P(t) \sim t^{-\tau_t}$ imply
that the average size of avalanches of duration $t$, $\bar{s}_t$, takes
the form
\begin{equation}
  \bar{s}_t \sim t^{m}, \qquad \qquad \mathrm{with} \qquad m =
  \frac{\tau_t -1}{\tau_s -1}.
  \label{eq:average_s_vs_t}
\end{equation}
In Fig.~\ref{fig:distributions_fish}(e) we represent the empirical
average avalanche size $\bar{s}_t$ as a function of the duration
$t$. For the different values of the turning threshold considered, we
estimate numerically that $\overline{s} \sim t^{1.31}$. This observation
is in good agreement with the expression in
Eq.~\eqref{eq:average_s_vs_t}, which, using the values from
Table~\ref{tab:exponents}, yields $m = 1.4(2)$.  In
Fig.~\ref{fig:distributions_fish}(e) we also show the average avalanche
size observed in randomized avalanches, which shows a linear dependence
as expected theoretically, see Methods. This last result highlights the
relevant effect of turning angle correlations in real fish.

\begin{table}[b]
  \centering

  \begin{tabular}{c|c|c|c|c|}
    \cline{2-5}
    \multicolumn{1}{l|}{} & \multicolumn{4}{c|}{Schooling fish} \\ \cline{2-5}
                          & $\tau_s$ & $\sigma_s$ & $\tau_t$ & $\sigma_t$ \\ \cline{2-5}
                          & 2.0(1) & 3.1(3) & 2.4(1) & 1.70(4) \\ \cline{2-5}
    \multicolumn{1}{l|}{} & \multicolumn{4}{c|}{\begin{tabular}[c]{@{}c@{}}Vicsek model with a perturbed leader\\ $\varphi_\mathrm{th}(\eta) = 2.5 \pi \eta$\end{tabular}} \\ \cline{2-5}
                          & $\tau_s$ & $D$ & $\tau_t$ & $z$ \\ \hline
    \multicolumn{1}{|c|}{$\eta=0.2$} & 1.73(5) & 2.01(2) & 4.03(5) & 0.39(5) \\ \hline
    \multicolumn{1}{|c|}{$\eta=0.3$} & 1.69(5) & 2.01(2) & 3.49(5) & 0.42(5) \\ \hline
    \multicolumn{1}{l|}{} & \multicolumn{4}{c|}{\begin{tabular}[c]{@{}c@{}}Vicsek model with a perturbed leader\\ $\varphi_\mathrm{th}(\eta) = 2.8 \pi \eta$\end{tabular}} \\ \cline{2-5}
    \multicolumn{1}{l|}{} & $\tau_s$ & $D$ & $\tau_t$ & $z$ \\ \hline
    \multicolumn{1}{|c|}{$\eta=0.2$} & 0.99(5) & 2.06(2) & 0.26(4) & 0.50(5) \\ \hline
    \multicolumn{1}{|c|}{$\eta=0.3$} & 1.04(5) & 2.03(2) & 0.57(5) & 0.52(5) \\ \hline
  \end{tabular}

  \caption{Summary of scaling exponents for the avalanche size and
    duration distributions obtained from observations of a real fish
    school and from the Vicsek model with a perturbed leader.}
  \label{tab:exponents}
\end{table}

\subsubsection{Effective leadership and avalanche behavior}

\begin{figure}[t]
  \centering{\includegraphics[width=\columnwidth]{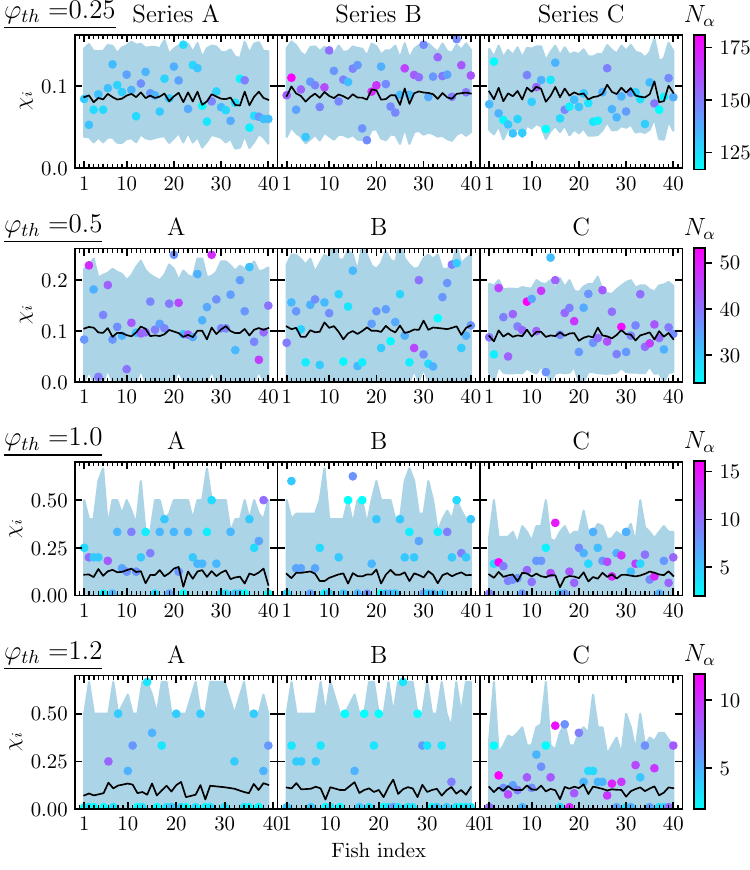}}

  \caption{\label{fig:leadership_fish} Leadership probability $\chi_i$
    computed for each fish $i$ in the three data series considered A, B
    and C  columns, left to right) and for different
    values of the turning threshold $\vphit=0.25, 0.5, 1.0$ and $1.2$
    (rows, top to bottom). Symbols are color-coded with
    the number $N_\alpha$ of actual avalanches in which each fish
    participates. Full lines represent the average leadership
    probability in a null model of uncorrelated avalanches. The shadowed
    regions represents the $99\%$ confidence interval of this value.}

\end{figure}

In order to explain the origin of the avalanche behavior observed in our
empirical data, we consider the possibility that avalanches are
triggered by some initiator or effective \emph{leader}, which
consistently starts the large turning rearrangements that lead to the
formation of an avalanche. While several definitions of leadership have
been proposed within the field of collective animal
motion~\cite{Strandburg-Peshkin.2018}, here we use a measure explicitly
devised to detect the presence of preferential initiators of avalanches.
We consider the originator of an avalanche as the fish that performs the
first large heading turn in the evolution of the avalanche. As more than
one fish can be active in any frame, we consider as initiators all
active fish in the first frame of an avalanche.  We define the
\emph{leadership probability} $\chi_i$ of fish $i$ in a given data
series as the ratio of the number of avalanches in which the fish $i$ is
active in the first frame, divided by the total number of avalanches in
which fish $i$ participates.  The calculation is restricted to
sufficiently large avalanches, of duration larger than $5$ frames. In
Fig.~\ref{fig:leadership_fish} we plot the value of $\chi_i$ computed
for each one of the $N=40$ fish in each series, for different values of
the turning threshold $\vphit$. As we can see, the leadership
probability shows an important variation among fish.  Moreover, for the
largest values of $\vphit$ considered, the leadership probability can
take values up to $0.60$, indicating that some fish initiate more than
half of the avalanches in which they participate.

In order to quantify the relevance of the values of $\chi_i$ obtained,
and ascertain that they are not the effect of random fluctuations in the
activity of the fish, given our small populations, we compare our
empirical estimates with the results obtained in a null model in which
the turns performed by fish are completely independent, see Methods.
The continuous lines in Fig.~\ref{fig:leadership_fish} represent the
null model average leadership probability, while the shaded region
represents its $99\%$ confidence interval.  Our results in
Fig.~\ref{fig:leadership_fish} indicate that, with the exception of
series C, in all series and for all values of the turning threshold
considered, several fish have an unusually very large probability to
initiate an avalanche, much larger than the value expected from pure
random fluctuations. We can associate them to effective leaders of the
school, which initiate with large probability the avalanches in which
they participate.

More information can be obtained by considering the evolution of the
leadership probability as a function of the turning threshold for each
fish in each time series, see Supplementary Figure (SF)~1. From this
plot we can confirm, first of all, that some fish never initiate an
avalanche ($\chi_i = 0$) for large values of the turning threshold,
while others consistently start much less avalanches than they should by
mere random fluctuations. Some other fish behave as initiators for some
range of values of the turning threshold.  Finally, some fish reliably
initiate a large number of avalanches, much more than they should by
pure randomness. These fish can be identified as consistent effective
leaders, which trigger a large majority of the avalanches, independently
of the value of the threshold used to quantitatively define them.

\subsection{Modeling avalanches in the presence of leaders}

To explore the effects of the presence of leadership in the avalanche
behavior of schooling fish, we consider as the simplest modeling
scenario a variation of the classic Vicsek model~\cite{vicsek1995} in
which we introduce an effective leader.

\subsubsection{Model definition}

The Vicsek model~\cite{vicsek1995,Ginelli2016} is defined in terms of
$N$ self-propelled particles (SPPs), characterized by a position
$\vec{r}_i$ and a velocity $\vec{v}_i$ of constant modulus $v_0$,
evolving in a two dimensional space, and thus being fully characterized
by the heading angle $\theta_i$ defined by the velocity
vector. Particles interact among them by trying to align their
instantaneous velocity with the average velocity of the set of nearest
neighbors inside a circular region of radius $R$ centered in the
considered SPP. A noise source of strength $\eta$, representing physical
or cognitive difficulties in gathering or processing local information,
allows the formation of an ordered (\emph{flocking}) phase at low noise
intensities, and of disordered (\emph{swarming}) states at high enough
noise values. See Methods for further details and simulation conditions.

In the variation we study (see Methods) we consider that a given
particle, say particle $1$, plays the role of an effective global leader
with a long range influence over the rest of the SPPs. We notice
  that long-range interactions have already been considered in models of
  collective motion, as a mechanism to ensure a compact flock in the
  absence of periodic boundary conditions\cite{Zumaya.2018}.The
velocity of the leader, $\vec{v}_1(t) = \vec{v}_L$ is not affected by
the behavior of its neighbors, and therefore its heading could remain
constant $\theta_1(t) = \theta_L$ over time.  The other SPPs can, on the
other hand, feel the orienting effect of their local neighborhood as
well as that of the leader, independently of their relative
distance. Therefore they take it into account when computing the average
velocity of their neighbors, to which they try to align.

At this point, it is worth pointing out that this leader can be any
individual who first experiences a sudden orientational shift. For this
reason we simply assume that its heading will remain constant, and
therefore unaffected by
its neighbors, until another reorientation of
similar characteristics occurs in the system. We have checked that the
leader
does not need to be always the same individual to obtain the
main results of our model. On the other hand, the presence of an
unperturbed leader, that is, a leader with a constant heading over time,
would have the effect of suppressing the disordered phase exhibited by
the classic Vicsek model. As we can see in SF~2, while for the classical
model the transition becomes sharper when increasing the systems size
$L$, the leader induces an ordered state for any value of $\eta$, with
an order parameter (see Methods) fairly independent of system size and
vanishing only in the limit of maximum noise $\eta=1$.

\subsubsection{Avalanche behavior in response of leader perturbations}
\label{sec:avalanche-behavior}

In this section, we focus our attention on the system-wide perturbations
that are induced by changes in the preferred direction of motion of the
leader. To analyze them, we consider a random reorientation of the
leader's heading by an angle $\Delta \theta_L$, performed in the steady
state corresponding to a given value of the noise intensity $\eta$, and
measure the subsequent rearrangements that this perturbation induces in
the heading of the rest of fish, as given by the turning angle
$\varphi_i(t) = \theta_i(t+1) - \theta_i(t)$ projected on the interval
$[0, \pi]$.

\begin{figure}[t]
  \centering{\includegraphics[width=\columnwidth]{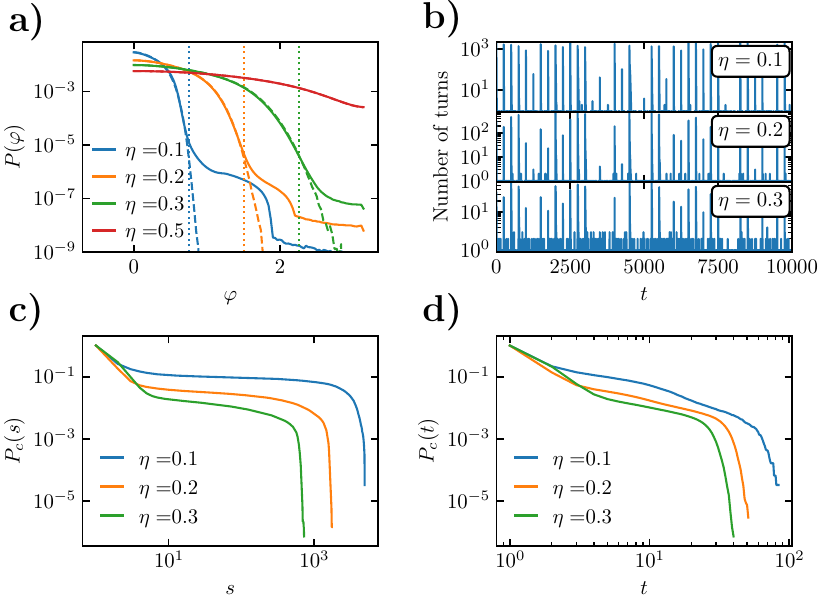}}

  \caption{\label{fig:vicsek_stats} (a) Probability density of the
    turning angles $\varphi$ for the Vicsek model with a leader, for
    different values of the noise intensity $\eta$, in a system of size
    $L=220$. Dashed lines correspond to a non-turning leader. Full lines
    represent a leader perturbed periodically every $250$ time steps.
    Vertical lines, indicating the departure of the distributions for
    perturbed and non-perturbed leaders, are estimated at a value
    $\varphi_c(\eta) \simeq 2.4 \pi \eta$ (b) Number of SPPs turning an
    angle larger than $\vphit(\eta) = 2.5 \pi\eta$ in a sequence of
    $10000$ simulation time steps in the Vicsek model with a
    periodically perturbed leader, for different values of $\eta$. (c)
    Cumulated distribution of sizes $P_c(s)$ of avalanches induced by a
    periodically perturbed leader in a system of size $L=220$ with
    turning threshold $\vphit(\eta)$, for different values of the noise
    intensity. (d) Cumulated distribution of durations $P_c(t)$ of
    avalanches induced by a periodically perturbed leader for different
    values of the noise intensity.}
\end{figure}

In Fig.~\ref{fig:vicsek_stats}(a) we represent the probability density
of SPPs turning angles $P(\varphi)$ in the steady state, for different
values of the noise intensity $\eta$. In this plot we consider the model
with a fixed, non-turning leader (dashed lines), and the case of a
periodically perturbed leader (full lines), in which the leader
experiences a random rotation $\Delta \theta_L$ of its heading,
uniformly distributed in the interval $[-\pi, \pi]$, every 250 time
steps, a time lapse larger than the maximum avalanche duration recorded
in simulations.  As we can see, for fixed $\eta$, the two distributions
are almost identical for small $\varphi$, while they differ drastically
regarding the behavior of the tails beyond a given cut-off turning angle
$\varphi_c(\eta)$.  A numerical analysis performed for different values
of $L$ allows to estimate this cut-off as
$\varphi_c(\eta) \simeq 2.4 \pi \eta$.  The presence of this turning
angle cut-off, not available in empirical data, permits to distinguish
the changes of heading due to the effect of the leader perturbations,
and suggests that the proper definition of avalanches should consider
turning thresholds larger than the cut-off $\varphi_c(\eta)$.  In the
following, we will fix the value of the threshold to
$\vphit(\eta) = 2.5 \pi \eta$. We notice that for large $\eta \geq 0.5$,
the angular distributions with and without perturbations are identical,
compatible with a large noise masking external perturbations and making
avalanches non discernible.

In Fig.~\ref{fig:vicsek_stats}(b) we plot a sample of the number of SPPs
that turn an angle larger than $\vphit(\eta)$ as a function of
time. This curve emphasizes the heterogeneous character of the avalanche
sizes in response to the leader's changes of direction, akin to what is
observed in fish schools: Sometimes a perturbation is followed by a
small number of SPPs reorientations; but other times, it triggers the
reorientation of a large number of particles.  As expected, the strength
of the effects of the leader perturbations decreases with increasing
noise, indicating that interesting avalanche behavior will only occur
for moderate levels of noise.

We compute the cumulative probability distributions $P_c(s)$ and
$P_c(t)$, Eq.~(\ref{eq:distrib}), of observing an avalanche of size and
duration larger than $s$ and $t$, respectively, plotted in
Fig.~\ref{fig:vicsek_stats}(c) and (d) for a turning threshold
$\vphit(\eta)$ and different values of $\eta$.  As we can see from these
plots, the values $\eta=0.2$ and $0.3$ lead to size an time
distributions analogous to that observed in rearrangement avalanches in
real fish schools, with a shape that can be approximated by a power-law
form for intermediate values, followed by a crossover to a sharp
decrease for large $s$ and $t$ above a characteristic size or time.  The
behavior for $\eta=0.1$ is more complex, probably due to the fact that
for small noise one expects a fairly homogeneous response with many SPPs
following a leader perturbation. We thus discard this value in the
following analysis.

\begin{figure}[t]
  \centering{\includegraphics[width=\columnwidth]{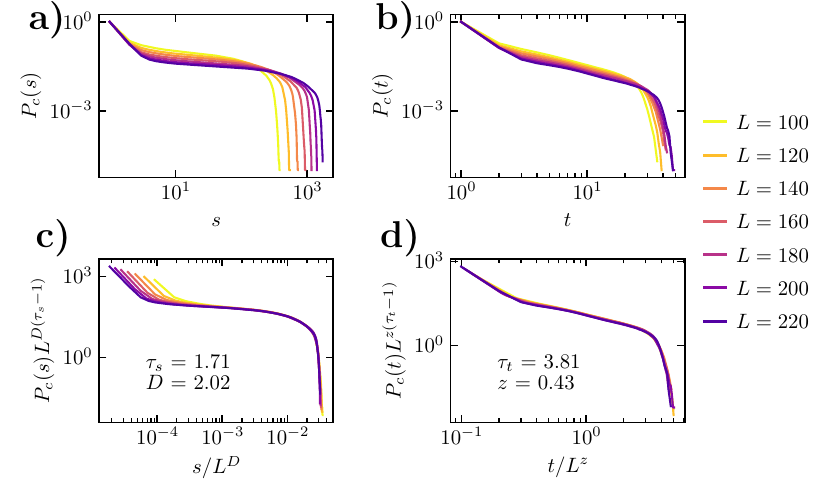}}

  \caption{\label{fig:vicsek_size_scaling} (a) Cumulative probability
    distribution of size $P_c(s)$ of avalanches induced by a perturbed
    leader in a system with $\eta=0.2$, turning threshold $\vphit(\eta)$
    and different values of $L$.  (b) Cumulative probability
    distribution of durations $P_c(t)$ of avalanches induced by a
    perturbed leader.  (c) Check of the scaling of the cumulated size
    distribution as given by Eq.~(\ref{eq:collapse}). (d) Check of the
    scaling of the cumulated time distribution as given by
    Eq.~(\ref{eq:collapse}).  Statistics are performed over at least
    $10^5$ different avalanches.}
\end{figure}

The fact that we work now with a numerical model, allows us to explore
the behavior of the system for different systems sizes $L$ at a fixed
turning threshold, which was not possible in our fixed size empirical
data. In Fig.~\ref{fig:vicsek_size_scaling}(a) and~(b) we plot the
cumulative size and duration distributions in avalanches in the Vicsek
model with a turning leader for a turning threshold $\vphit(\eta)$,
$\eta=0.2$ and different system sizes.  As we can observe, the behavior
of the distributions is analogous to that observed in real fish schools,
compatible with a power-law decay but that are now truncated by a size
and time cut-offs that are functions of the system size $L$. Inspired
again by self-organized criticality~\cite{pruessner2012}, we can assume
now that the distributions obey a finite-size scaling form
\begin{equation}
  \label{eq:3}
  P(s) = s^{-\tau_s} G_s\left( \frac{s}{L^{D}} \right), \quad
  P(t) = t^{-\tau_t} G_t\left( \frac{t}{L^z} \right).
\end{equation}
where $D$ and $z$ are new characteristic exponents that define the
characteristic size $s_c(L) \sim L^D$ and time $t_c(L) \sim L^z$ as a
function of the system size~\cite{pruessner2012,cardy88}. The better
statistics in numerical simulations allow to estimate the characteristic
exponents applying the more precise moments analysis
technique~\cite{PhysRevE.58.R2677}, see Methods. Application of this
method leads to the characteristics exponents reported in
Table~\ref{tab:exponents}. We check the accuracy of these values
performing a data collapse analogous to that performed for the
avalanches in real fish, which, for the cumulated distributions, takes
the form,
\begin{equation}
  \label{eq:6}
  L^{D(\tau_s - 1)} P_c(s) = F'_s(s L^D), \qquad  L^{z(\tau_t - 1)}
  P_c(t) = F'_t(t L^z),
\end{equation}
see Figs.~\ref{fig:vicsek_size_scaling}(c) and~(d) for $\eta=0.2$; the
case $\eta=0.3$ is presented in SF~3. As we can see from these values,
the exponents show a dependence on the value of the noise $\eta$,
although the size exponents appear to be compatible within error
bars. It is important to notice that the presence of a rotating leader
in necessary to obtain scaling avalanche distributions. Even in the
absence of a leader, the heading fluctuations due to noise and
interactions in the standard Vicsek model allow to define avalanches for
a given threshold. These avalanches, however, show a simple, short
ranged exponential distribution, as shown in SF~4.

We have finally checked the effects of changing the turning threshold in
the scaling of the distributions as a function of the system size. In
SF~5 and SF~6 we show the results for a turning threshold
$\vphit = 2.8 \pi \eta$, summarized in Table~\ref{tab:exponents}. As we
can see, the scaling exponent $\tau_s$ and $\tau_t$ in our model depend
on the value of the threshold. This fact is in contrast with the
behavior of the fish school, in which the characteristics exponents
appear to be independent of the threshold, and thus allowing for a
scaling solution of the form given by
Eq.~(\ref{eq:scaling}). Interestingly, the exponents $D$ and $z$ appear
to be rather detail independent, taking the approximate values
$D \simeq 2$ and $z \simeq 1/2$ for any value of $\eta$ and $\vphit$,
which would indicate that avalanches in this model are
compact~\cite{pruessner2012}.

\section{Discussion}

Behavioral cascades, taking the form of intermittent rearrangements
(avalanches) in the patterns of movement are an important, albeit
sometimes neglected, feature of collective motion in animals. Here we
have shown that behavioral cascades can be observed in the rearrangement
dynamics of swimming fish schools. Such avalanches, defined in terms of
a turning threshold for the heading of the fish, have distributions of
sizes and times exhibiting a scaling behavior compatible with a
power-law tail truncated by a cut-off that is an increasing function of
the turning threshold.  A data collapse analysis allows to determine the
exponents characterizing the scaling form. We conjecture that such
avalanche behavior can be due to the presence of effective leadership in
the schools.  In order to support this conjecture, we introduce a
measure of leadership, based in the concept of avalanche initiators, and
observe that, indeed, some fish have consistently an unusually large
probability to initiate any avalanche in which they participate. These
predominant initiators can be interpreted as effective leaders,
determining the start of sudden rearrangements of the school
headings. Leadership in the context of avalanche initiation could
account for individuals having sudden behavioral changes or specific
information about the environment, such as the proximity of a wall.

To check whether the presence of leaders is enough to induce avalanche
behavior in collective motion, we have considered a very simple model,
consisting in a variation of the classical Vicsek model with the
addition of a global leader, which influences the movement of all other
particles, subject to random changes in its heading. Interestingly, this
simple model displays an intermittent behavior qualitatively similar to
that observed in real schools, with avalanche size and duration
distributions displaying a self-similar scaling form.

Our results provide a new perspective on the avalanche behavior observed
in real collective motion situations~\cite{Ginelli2015}, which can be
associated to a simple mechanism of leadership that exerts a long range influence, observed in many natural
situations, indicating the
possibility of a direct relation between these
phenomena. Leadership in the present context of a moving school
corresponds to those individuals that first react to any external input,
or that first exhibit a random behavioral change, and preferentially
start sudden rearrangements of the trajectories of other fish in the
school. Such interpretation is validated by the numerical results from
our model. It is also worth emphasizing that, while it does not offer a
perfect quantitative prediction of the characteristic exponents, it
nevertheless allows to reproduce the scaling form of the avalanche
distributions within a minimalist modeling framework.

Different venues of future research stem from the results presented
here. From an empirical perspective, it would be interesting to further
study the nature of the avalanches observed in real schools, and to
correlate them with other physical properties measured in similar
systems~\cite{Herbert-Read.2011,katz2011}, as well as with other
measures of leadership devised in other contexts of collective
motion~\cite{Strandburg-Peshkin.2018}. From a numerical point of view,
our results present new challenges in the understanding of the
properties of the proposed model. Indeed, a clearly open question
remains to ascertain the ultimate origin of the scaling behavior
observed in avalanches in a system in which no apparent critical
transition exists. Another interesting question regards the effects of
leader switching strategies. We expect the scale-free nature of the
observed avalanches to be preserved, provided that the influence of the
leader, sensory wise, remains rather long-ranged.  In this sense, as we
have numerically checked (data not shown), a short-ranged leader, only
with local influence over its nearest neighbors, is not able to induce
system-wide orientation rearrangements.  On the other hand, the value of
the exponents associated to the size and duration cutoffs are apparently
independent of the noise intensity imposed on the system. These
observations hint towards a possible partial universality, which is not
shared, however, by the power-law decay exponent. Further work in this
direction is clearly needed in order to clarify these issues.

\section{Methods}

\subsection{A null model of fish avalanches}
\label{sec:null_model_avalanches}

In the absence of any sort of dynamical correlations between the turning
angles of fish, the evolution of avalanches is purely determined by the
independent turning probability $P(\varphi)$ of each fish. As a null
model of avalanche behavior, we consider the case in which each fish
independently turns a angle $\varphi$ at each time step. Consider an
avalanche of duration $t$ and size $s$, starting at time $t'=1$. If the
avalanche lasts $t$ time steps, it means that at least one fish turned
an angle larger than $\vphit$ every frame from $t'=1$ to $t'=t$, and
that no fish turned an angle larger than $\vphit$ at frame $t' =
t+1$. Under these conditions, the probability that a fish turns an angle
larger than $\vphit$ in any frame is
\begin{equation}
  q = \int_{\vphit}^{\pi} P(\varphi) \; d\varphi,
  \label{eq:sup_turningprobn}
\end{equation}
and the probability that at least one fish turns an angle larger than
$\vphit$ in a given frame is
\begin{equation}
  Q = 1 - (1-q)^N,
\end{equation}
where $N$ is the number of fish. Therefore, the normalized probability
that an avalanche lasts for $t$ frames in this null model is
\begin{equation}
  P_0(t) = \frac{( 1- Q ) Q^t}{\sum_{t' =1}^\infty ( 1- Q ) Q^{t' }} = (1-Q)
  Q^{t-1}, \; t = 1, 2. \ldots, \infty,
  \label{eq:sup_timedist}
\end{equation}
where we consider that avalanches have a minimal duration of one frame.
That is, in the uncorrelated null model, the avalanche duration
distribution has an exponential form, with average avalanche duration
$\av{t}_0 = \sum_{t=1}^\infty t P_0(t) = 1 / (1-Q)$

Consider now a frame in an avalanche of finite duration. In this frame,
at least one fish turned an angle larger than $\vphit$, therefore the
the probability of observing the $s_1$ large turns in this frame is
\begin{equation}
  p_1(s_1) = \frac{1}{Q}\binom{N}{s_1} q^{s_1} (1-q)^{N -s_1},
  \; s_1 = 1,  2, \ldots, N,
  \label{eq:sup_oneframe}
\end{equation}
If the avalanche has duration $t$, at each frame a number $s_1$ of fish,
distributed with the probability Eq.~(\ref{eq:sup_oneframe}), will turn
a large angle. Therefore, the distribution of sizes in avalanches of
duration $t$, $P_0(s | t)$ will be given the convolution of the
probability Eq.~(\ref{eq:sup_oneframe}) $t$ times with itself. The form
of this expression is hard to compute. However, we can approximate
the avalanche size distribution as follows: Since
Eq.~(\ref{eq:sup_oneframe}) is similar to a binomial distribution, it is
bell-shaped and centered at the average value
\begin{equation}
  \bar{s}_1 = \sum_{s_1=1}^N s_1 p_1(s_1) = \frac{Nq}{Q}.
  \label{eq:mean_size_randomized}
\end{equation}
Therefore, the average size of an avalanche of duration $t$ is
\begin{equation}
  \bar{s}_t = \frac{Nq}{Q} t,
\end{equation}
linear with $t$. Assuming that the relation between size $s$ and
duration $t$ is tight, given the bounded distribution $p_1(s_1)$, we can
use relation $s \simeq \frac{Nq}{Q} t$ and the distribution $P_0(t)$
from Eq.~\eqref{eq:sup_timedist} to obtain the probability
transformation $P_0(t) dt = P_0(s)ds$, leading to
\begin{equation}
  P_0(s) \simeq \frac{1-Q}{Nq} e^{- s Q \ln(1/Q)/(N q)},
\end{equation}
that is, an exponential decay with a characteristic size
\begin{equation}
  s_c = \frac{Nq}{Q \ln(1/Q)}.
\end{equation}

\subsection{Numerical data collapse analysis}

We start from a set of avalanche size (or duration) distributions, that we assume to
fulfill the scaling relation, at the level of the cumulated distributions,
\begin{equation}
  P_c(s) = s^{-\tau_s + 1} F_s\left( s \vphit^{\sigma_s}\right ).
\end{equation}
In order to find the exponents $\tau_s$ and $\sigma_s$, we proceed as follows: We
consider general exponents $x_s$ and $y_s$, from which we can write the new rescaled
expressions
\begin{equation}
\vphit^{y_s ( 1 - x_s )} P_c(s) = F'_s\left( s \vphit^{y_s}\right ),
\end{equation}
where $F'_s(z) = z^{-\tau_s +1} F_s(z)$. Plotting
$\vphit^{y_s ( 1 - x_s )} P_c(s)$ as a function of $s \vphit^{y_s}$, the
curves for different values of $\vphit$ will collapse onto the universal
function $F'_s(z)$ when $x_s = \tau_s$ and $y_s = \sigma_s$. We can
estimate this exponent by considering the difference of the curves for
the different values of $\vphit$ and choosing the exponents $\tau_s$ and
$\sigma_s$ as the values of the exponents $x_s$ and $y_s$ that minimize
this difference. To compute the difference, we locate the interval of
values of $s \vphit^{y_s}$ common for all $\vphit$. In this interval, we
compute a spline of order $k$ for each quantity
$\vphit^{y_s ( 1 - x_s)} P_c(s)$ and interpolate a fixed number $n$ of
equispaced points. The difference is defined as the sum of the variances
of the values of $\vphit^{y_s ( 1 - x_s )} P_c(s)$ in each point of the
interpolation, for the different values of $\vphit$.  In the results
presented here, we consider splines of order $k=2$ and interpolate
$n=10$ points for each $P_c(s)$ curve.

\subsection{Leadership probability in the null model of avalanche
  behavior}

In the avalanche null model defined above, consider a fish that
participates in a given avalanche. To estimate its leadership
probability we have to compute the probability that it leads the
avalanche (i.e.\ it is active in its first time step), provided that it
participates in it. To compute it, we use Bayes' theorem to write
\begin{equation}
  P(p) P(l | p) = P(l) P(p | l),
  \label{eq:sup_bayes}
\end{equation}
where $P(p)$ is the probability that the fish participates in a given
avalanche, $P(l | p)$ the probability of leading an avalanche in which
it participates (the probability we are seeking), $P(l)$ the probability
of leading an avalanche, and $P(p | l)$ the probability that a fish
participates in an avalanche provided that it leads it.  Obviously,
$P(p | l) = 1$. To estimate the rest of probabilities, we need
information about the duration $t$ of the avalanche. Thus, we have
$P(p) = 1 - (1 - q)^{t}$, the probability that the fish turns at least
once in the development of the avalanche, and $P(l) = q$, the
probability that the fish is active (performs a large turn) in the first
time step of the avalanche. Therefore, from Eq.~\eqref{eq:sup_bayes} we
obtain
\begin{equation}
  P(l | p) = \frac{P(l) P(p | l)}{P(p)} = \frac{q}{1 - (1-q)^t}.
\end{equation}

Within this null model, consider a fish that participates in $N_a$
avalanches, each of duration $t_\alpha$, $\alpha = 1, \ldots, N_a$. The
probability of leading any of these avalanches is
$p_\alpha = q / [1 - (1-q)^{t_\alpha}]$. Therefore, the probability
$P(\ell)$ of leading $\ell$ of the $N_a$ avalanches is given by a
Poisson binomial distribution, representing the probability distribution
of a sum of independent Bernoulli trials that have different success
probabilities $p_\alpha$~\cite{10.1214/aoms/1177728178}.  The Poisson
binomial distribution has a rather convoluted form, but its mean and
variance can be easily expressed as
\begin{eqnarray}
  \mu &=& \sum_\ell \ell P(\ell) = \sum_\alpha p_\alpha, \\
  \sigma^2 &=&  \sum_\ell \ell^2 P(\ell) - \left[  \sum_\ell \ell P(\ell)
  \right]^2 = \sum_\alpha p_\alpha(1-p_\alpha).
\end{eqnarray}
The average leadership probability of a fish in this null model is thus
given by $\chi_0 = \av{\ell} / N_a$, where
$\av{\ell} = \sum_\alpha p_\alpha$ is the average number of avalanches
led by the fish.  In Fig.~\ref{fig:leadership_fish} we show the actual
values of $\chi_i$ computed for each fish. The full line and shaded
region represents the null-model average value $\chi_i^0$ computed for
each fish, taking into account the number of avalanches in which it
participates, and its $99\%$ confidence interval, respectively.

\subsection{Vicsek model with leadership}

In the classic Vicsek model~\cite{vicsek1995,Ginelli2016}, $N$ self-propelled
particles (SPPs) move in a two dimensional space. The dynamics is overdamped and
defined in discrete time, with the instantaneous position $\vec{r}_i(t)$, $i=1,
\ldots, N$, of each particle being related with its velocity $\vec{v}_i(t)$ by
\begin{equation}
  \vec{r}_i(t+ \Delta t) = \vec{r}_i(t) + \vec{v}_i(t) \Delta t,
  \label{eq1}
\end{equation}
where $\Delta t$ is an integration time step, arbitrarily fixed to
$\Delta t = 1$.  Velocities have a constant modulus,
$|\vec{v}_i(t)| = v_0$, and thus are fully determined by their
direction, given by the heading angle $\theta_i(t)$ that the velocity
forms with, say the $x$ axis, such that
$\vec{v}_i(t) = (v_0 \cos \theta_i(t), v_0 \sin \theta_i(t))$.  Heading
is assumed to belong to the interval $[-\pi, \pi]$.  In this model each
particle $i$ tends to orient its direction of motion along the average
direction $\vec{V}_i(t)$ of the particles located inside a circular area
$\mathcal{V}_i$ of radius $R$ centered at its own position and including
itself, i.e.
\begin{equation}
  \label{eq:5}
  \vec{V}_i(t) = \frac{1}{n_i(t)}\sum_{j \in \mathcal{V}_i} \vec{v}_j (t),
\end{equation}
where $n_i(t)$ is the number of particles in the neighborhood
$\mathcal{V}_i$ at time $t$.
This dynamics is implemented in the update rule for the heading angle
\begin{equation}
  \label{eq:2}
  \theta_i(t +\Delta t) = \Theta \left[ \mathbf{V}_i(t) \right] + \eta
  \;  \xi_i(t),
\end{equation}
where the function $\Theta[\vec{V}]$ represents the angle of vector
$\vec{V}$ and $\xi_i(t)$ a random noise, uniformly distributed in the
interval $[-\pi, \pi]$, and $\eta \in [0,1]$ a parameter gauging the
strength of the noise term.

This model exhibits an order-disorder phase transition defined in terms
of an order parameter (polarization) given by
\begin{equation}
  \label{eq:1}
  \phi(\eta) = \frac{1}{v_0 N}\left \langle \left |  \sum_{i=1}^N
    \mathbf{v}_i(t)  \right| \right
    \rangle_t,
\end{equation}
the brackets representing a temporal average. The transition separates
an ordered phase for noise strength smaller than a critical value
$\eta_c$, corresponding to a flocking (schooling) phase, from a
disordered phase for $\eta > \eta_c$, corresponding to a swarming,
disordered phase.

In the variation of the Vicsek model we consider, one of the SPPs, say
particle $1$ plays the role of a leader which influences the orientation
of the rest of SPPs in the system, independently of their relative
distance. Therefore, in the heading update rule Eq.~\eqref{eq:2}, the
average velocity of the neighbors is replaced by the average
$\vec{V}_i^L(t)$ computed in the set
$\mathcal{V}_i^{L} = \mathcal{V}_i \; \cup \; \{ 1 \}$, including the
global leader and all the particles in the local neighborhood of $i$.
The heading and velocity of this leader is constant in time,
$\theta_1(t) = \theta_L$, and it represents a privileged direction it
wants to follow.

Simulations of the model are performed in square boxes of different size
$L$ with periodic boundary conditions. We fix the
density of particles $\rho = N / L^2 = 1$, the radius of interaction
$R=1$, and the constant speed of the SPPs $v_0 = 0.03$.

\subsection{Moments analysis technique}

The finite-size scaling (FSS) method~\cite{cardy88} assumes that the
dependence on system size $L$ of the avalanche size and time distributions
is of the form
\begin{eqnarray}
  \label{eq:4}
  P(s, L) & = & s^{-\tau_s} \mathcal{F}_s\left( \frac{s}{s_c(L)}\right), \\
  \label{eq:7}
  P(t, L) & = & t^{-\tau_t} \mathcal{F}_t\left( \frac{t}{t_c(L)}\right) ,
\end{eqnarray}
where $\mathcal{F}_x(z)$ are scaling functions that are approximately
constant for $z < 1$, and decay very fast to zero for $z>1$.  The quantities
$s_c(L)$ and $t_c(L)$ are the cut-off characteristics size and time, which
are assume to depend on system size as $s_c(L) \sim L^D$ and $t_c(L) \sim
L^z$, thus defining the standard critical exponents $\tau_s$, $\tau_t$, $D$
(the fractal dimension) and $z$ (the dynamic critical
exponent)~\cite{pruessner2012}.

Assuming the scaling form given by Eqs.~\eqref{eq:4} and~\eqref{eq:7}, we
can compute numerically the associated critical exponents applying the
moment analysis technique~\cite{PhysRevE.58.R2677}. One starts by defining
the $q$-th moment of the avalanche size distribution on a box of size $L$ as
\begin{eqnarray}
  \label{eq:8}
  \av{s^q}_L
    &=& \sum_s s^q\; P(s, L) \simeq \int ds\; s^{-\tau_s + q}
    \mathcal{F}_s\left( \frac{s}{L^D}\right) \nonumber \\
    &=& L^{D(q+1-\tau_s)} \int dx\; y^{-\tau_s + q} \mathcal{F}_s(x) \sim
      L^{\sigma_s(q)},
\end{eqnarray}
where we have introduce the FSS form in Eq.~\eqref{eq:4}, and taken the
continuous approximation for the $s$ and $t$ variables. The exponents
$\sigma_s(q) \equiv D(q+1-\tau_s)$ can be estimated as the slope of the
numerical evaluation of $\av{s^q}_L$ as a function of $L$ in a double
logarithmic plot. Then, for sufficiently large values of $q$, we can perform
a linear fit of the exponent $\sigma_s(q)$ to the form
\begin{equation}
  \label{eq:9}
  \sigma_s(q) = A q + B,
\end{equation}
with $A = D$ and $B = D(1-\tau_s)$, from where $D$ and $\tau_s$ can be
directly estimated. Along the same lines, the exponents associated to
the avalanche time distribution can be evaluated considering the $q$-th
moment of the time  distribution, $\av{t^q}_L \sim L^{\sigma_t(q)}$,
with $\sigma_t(q) \equiv z(q+1-\tau_t)$.

\section*{Acknowledgments}

We acknowledge financial support from the Spanish
MCIN/AEI/10.13039/501100011033, under Projects No.  PID2019-106290GB-C21
and No. PID2019-106290GB-C22. We thank F. S. Beltr\'an and V. Quera,
from the Institute of Neurosciences, University of Barcelona (Spain),
for providing the raw data of fish recordings.

\clearpage

\onecolumngrid
\renewcommand{\thefigure}{SM-\arabic{figure}}

\setcounter{figure}{0}
\section*{Supplementary Information}

\begin{figure}[t]

  \centerline{\includegraphics[width=0.9\columnwidth]{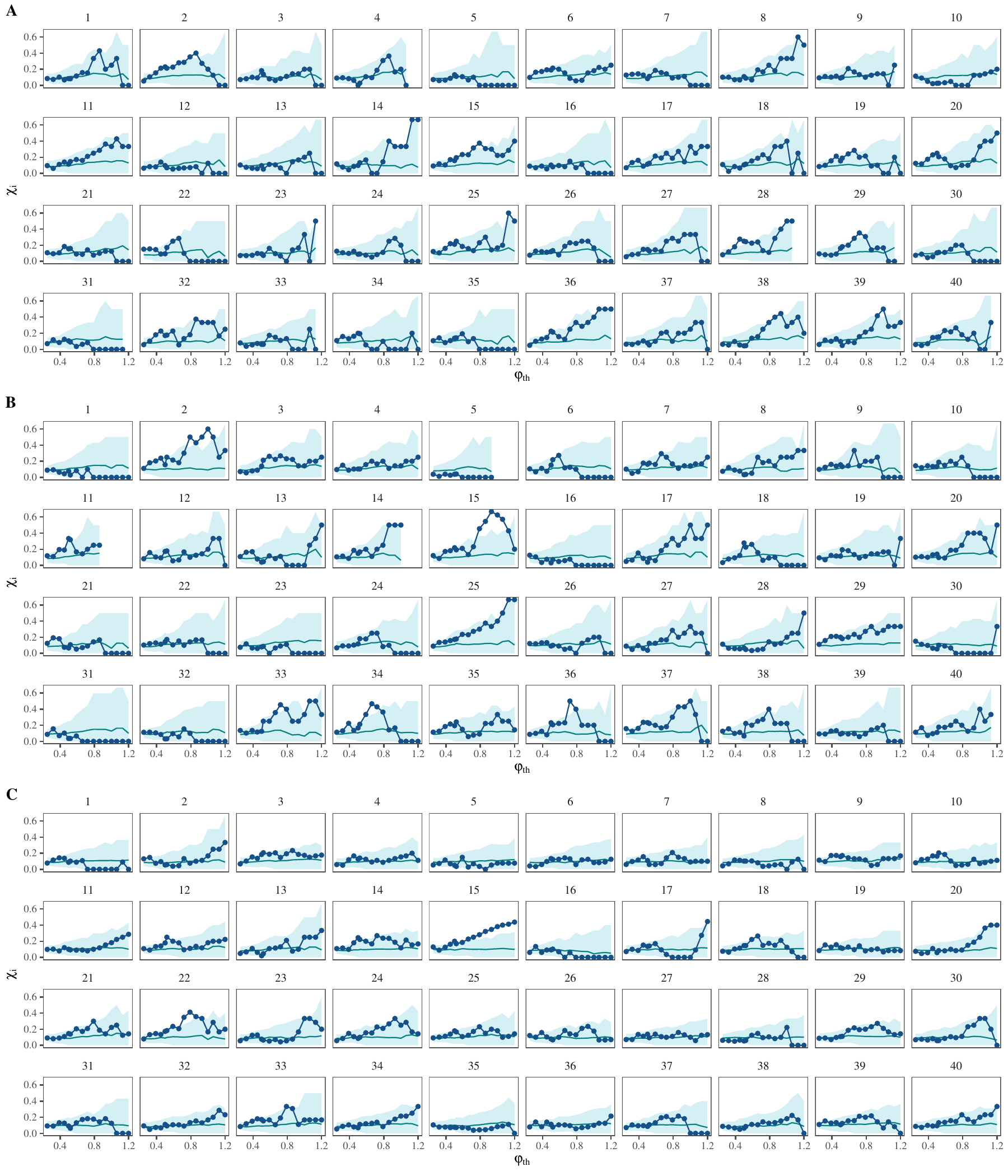}}

  \caption{Plot of the leadership probability $\chi_i$ as a function of
    the turning threshold $\vphit$, for the different fish in each
    series A, B, C (top to bottom).  Full lines represent the average
    leadership probability in a null model of uncorrelated
    avalanches. The shadowed regions represents the $99\%$ confidence
    interval of this value. Notice that in some plots certain values of
    $\vphit$ are missing. This is due to the fact that the corresponding
    fish do not participate in any avalanche, and therefore its
    leadership probability is not defined.}
  \label{fig:leadership_prob_evolution}
\end{figure}

\clearpage

\begin{figure}[p]
  \includegraphics[width=0.7\columnwidth]{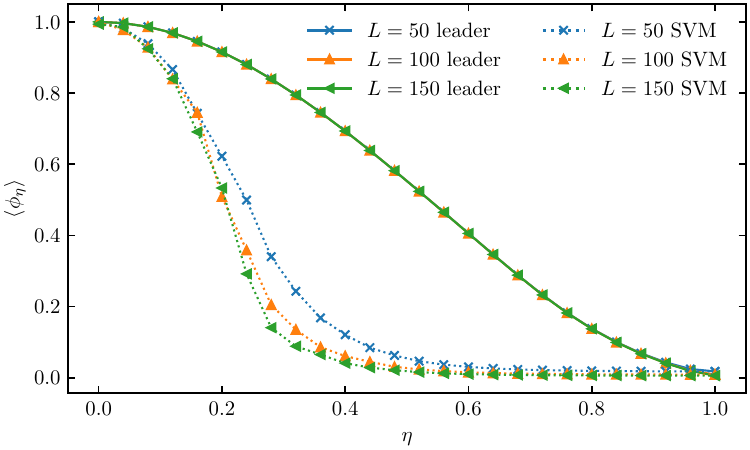} \caption{Average
    order parameter $\av{\phi(\eta)}$ as a function of noise intensity
    in the classic standard Vicsek model (SVM) and the Vicsek model with
    a non-rotating global leader for different system sizes.}
  \label{fig:sup_no_transition}
\end{figure}

\clearpage

\begin{figure}[t]
  \centering \includegraphics[width=0.7\columnwidth]{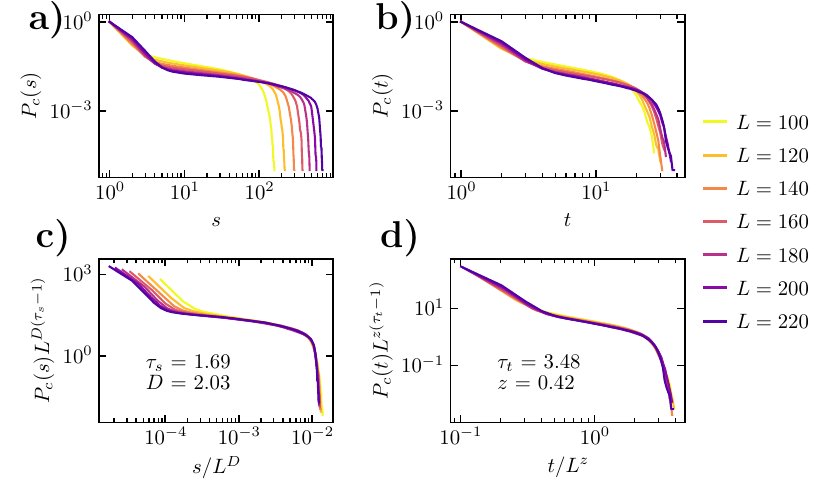}

  \caption{\label{fig:vicsek_size_scaling} (a) Cumulative probability
    distribution of size $P_c(s)$ of avalanches induced by a rotating
    leader in a system with $\eta=0.3$, turning threshold $\vphit(\eta)$
    and different values of $L$.  (b) Cumulative probability
    distribution of durations $P_c(t)$ of avalanches induced by a
    rotating leader in a system with $\eta=0.2$, turning threshold
    $\vphit(\eta)$ and different values of $L$. In both cases,
    statistics is performed over at least $10^5$ different
    avalanches. (c) Check of the scaling of the cumulated size
    distribution with turning threshold $\vphit(\eta)$, as given by
    Eq.~(8) in the main paper. (d) Check of the scaling of the cumulated
    time distribution with turning threshold $\vphit(\eta)$, as given by
    Eq.~(8) in the main paper.}
\end{figure}

\begin{figure}[t]
  \centering \includegraphics[width=0.7\columnwidth]{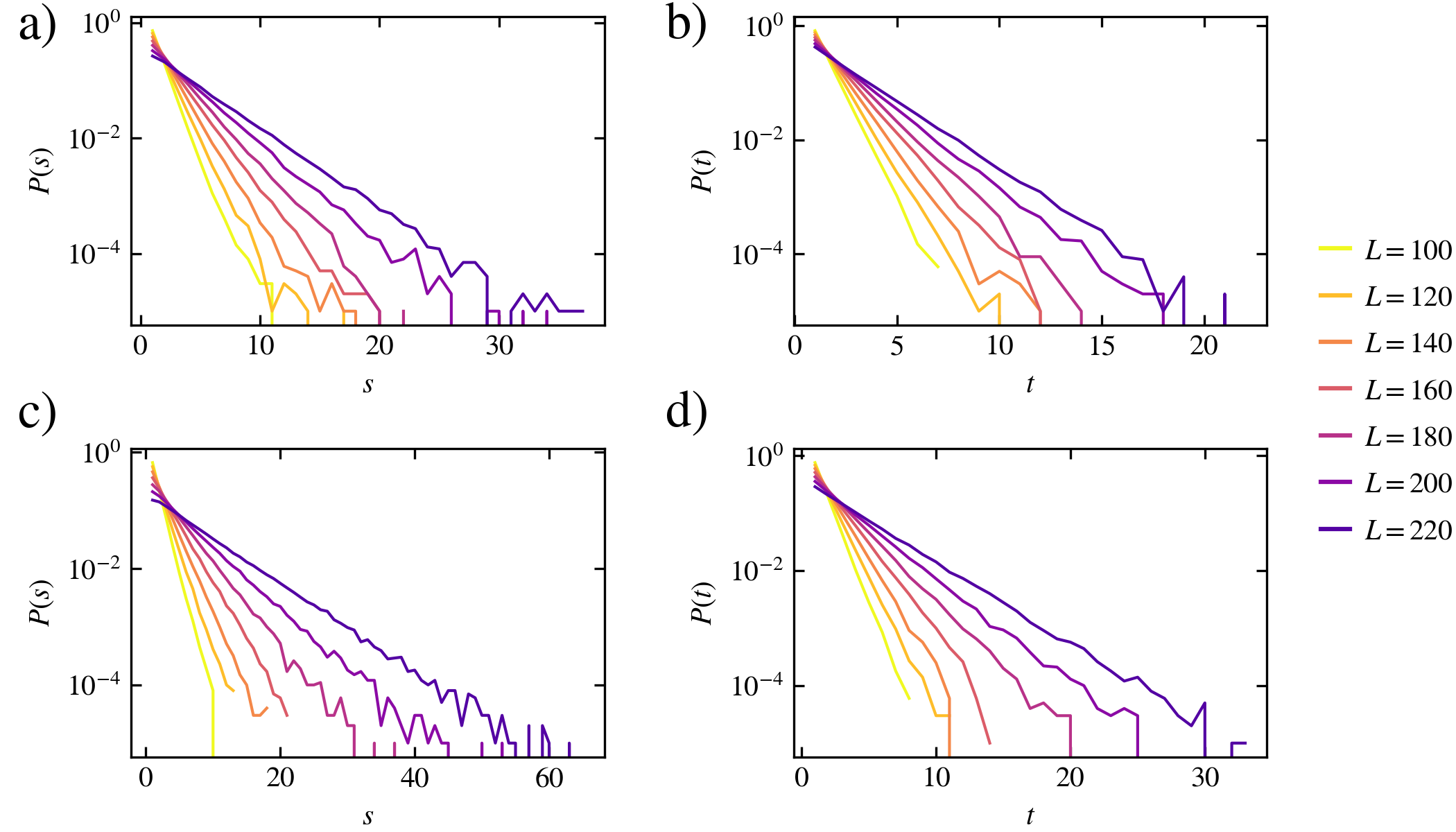}

  \caption{\label{fig:vicsek_size_scaling} (a) Cumulative probability
    distribution of size $P_c(s)$ of avalanches in the standard Vicsek model
    with $\eta=0.2$, turning threshold $\vphit(\eta)$ and different values of
    $L$.  (b) Cumulative probability distribution of duration $P_c(t)$ in the
    standard Vicsek model with $\eta=0.2$, turning threshold $\vphit(\eta)$ and
    different values of $L$.  (c) Cumulative probability distribution of size
    $P_c(s)$ of avalanches in the standard Vicsek model with $\eta=0.3$, turning
    threshold $\vphit(\eta)$ and different values of $L$.  (d) Cumulative
    probability distribution of duration $P_c(t)$ in the standard Vicsek model
    with $\eta=0.3$, turning threshold $\vphit(\eta)$ and different values of
    $L$.  Statistics are performed over at least $10^5$ different avalanches.}
\end{figure}

\begin{figure}[t]
  \centering \includegraphics[width=0.7\columnwidth]{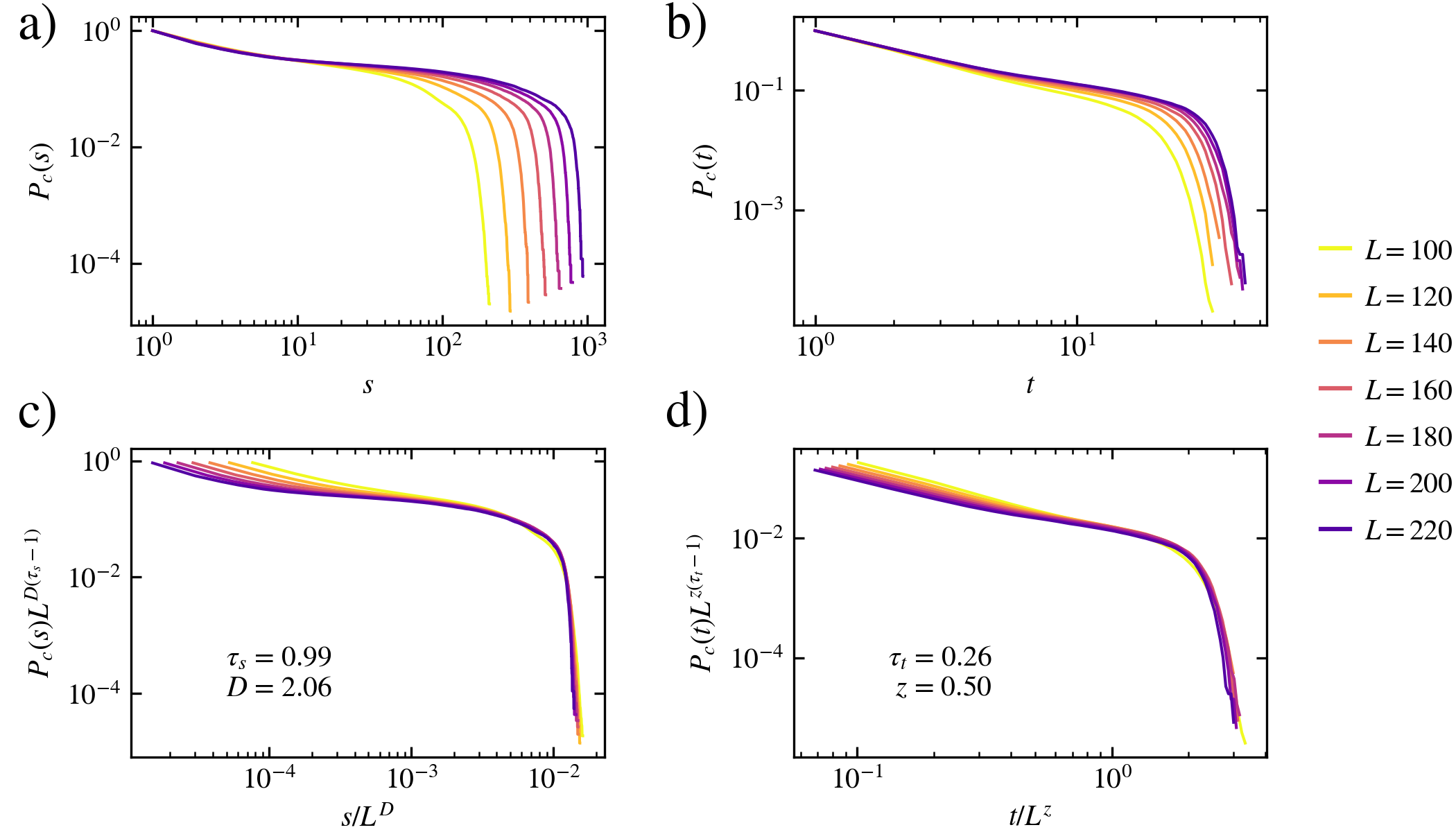}

  \caption{\label{fig:vicsek_size_scaling} (a) Cumulative probability
    distribution of size $P_c(s)$ of avalanches induced by a rotating
    leader in a system with $\eta=0.2$, turning threshold
    $\vphit = 2.8 \pi \eta$ and different values of $L$.  (b) Cumulative
    probability distribution of durations $P_c(t)$ of avalanches induced
    by a rotating leader in a system with $\eta=0.2$, turning threshold
    $\vphit(\eta)$ and different values of $L$. In both cases,
    statistics is performed over at least $10^5$ different
    avalanches. (c) Check of the scaling of the cumulated size
    distribution with turning threshold $\vphit(\eta)$, as given by
    Eq.~(8) in the main paper. (d) Check of the scaling of the cumulated
    time distribution with turning threshold $\vphit(\eta)$, as given by
    Eq.~(8) in the main paper.}
\end{figure}

\begin{figure}[t]
  \centering \includegraphics[width=0.7\columnwidth]{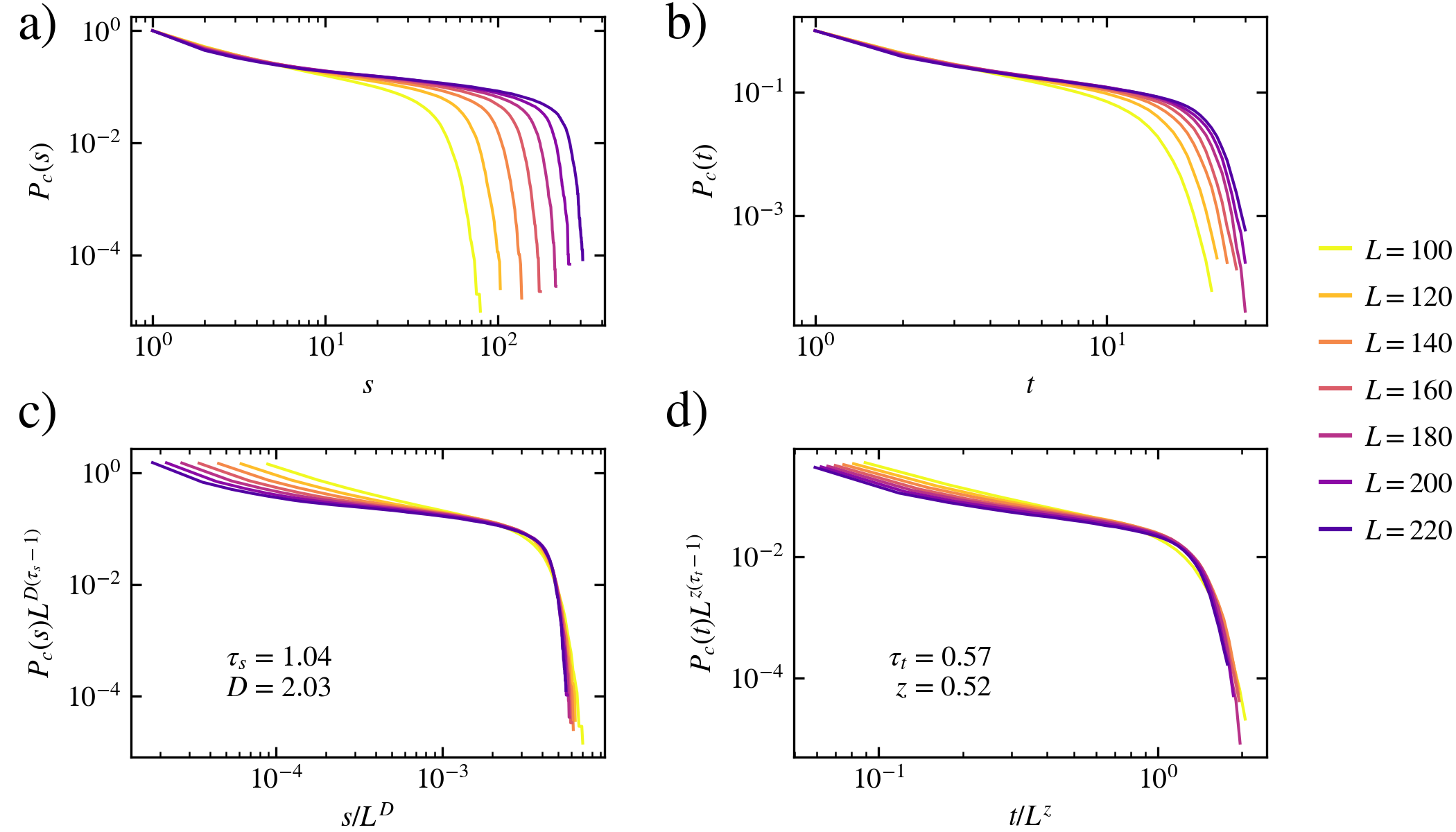}

  \caption{\label{fig:vicsek_size_scaling} (a) Cumulative probability
    distribution of size $P_c(s)$ of avalanches induced by a rotating
    leader in a system with $\eta=0.3$, turning threshold
    $\vphit = 2.8 \pi \eta$ and different values of $L$.  (b) Cumulative
    probability distribution of durations $P_c(t)$ of avalanches induced
    by a rotating leader in a system with $\eta=0.3$, turning threshold
    $\vphit(\eta)$ and different values of $L$. In both cases,
    statistics is performed over at least $10^5$ different
    avalanches. (c) Check of the scaling of the cumulated size
    distribution with turning threshold $\vphit(\eta)$, as given by
    Eq.~(8) in the main paper. (d) Check of the scaling of the cumulated
    time distribution with turning threshold $\vphit(\eta)$, as given by
    Eq.~(8) in the main paper.}
\end{figure}


\begin{thebibliography}{45}%
\makeatletter
\providecommand \@ifxundefined [1]{%
 \@ifx{#1\undefined}
}%
\providecommand \@ifnum [1]{%
 \ifnum #1\expandafter \@firstoftwo
 \else \expandafter \@secondoftwo
 \fi
}%
\providecommand \@ifx [1]{%
 \ifx #1\expandafter \@firstoftwo
 \else \expandafter \@secondoftwo
 \fi
}%
\providecommand \natexlab [1]{#1}%
\providecommand \enquote  [1]{``#1''}%
\providecommand \bibnamefont  [1]{#1}%
\providecommand \bibfnamefont [1]{#1}%
\providecommand \citenamefont [1]{#1}%
\providecommand \href@noop [0]{\@secondoftwo}%
\providecommand \href [0]{\begingroup \@sanitize@url \@href}%
\providecommand \@href[1]{\@@startlink{#1}\@@href}%
\providecommand \@@href[1]{\endgroup#1\@@endlink}%
\providecommand \@sanitize@url [0]{\catcode `\\12\catcode `\$12\catcode
  `\&12\catcode `\#12\catcode `\^12\catcode `\_12\catcode `\%12\relax}%
\providecommand \@@startlink[1]{}%
\providecommand \@@endlink[0]{}%
\providecommand \url  [0]{\begingroup\@sanitize@url \@url }%
\providecommand \@url [1]{\endgroup\@href {#1}{\urlprefix }}%
\providecommand \urlprefix  [0]{URL }%
\providecommand \Eprint [0]{\href }%
\providecommand \doibase [0]{http://dx.doi.org/}%
\providecommand \selectlanguage [0]{\@gobble}%
\providecommand \bibinfo  [0]{\@secondoftwo}%
\providecommand \bibfield  [0]{\@secondoftwo}%
\providecommand \translation [1]{[#1]}%
\providecommand \BibitemOpen [0]{}%
\providecommand \bibitemStop [0]{}%
\providecommand \bibitemNoStop [0]{.\EOS\space}%
\providecommand \EOS [0]{\spacefactor3000\relax}%
\providecommand \BibitemShut  [1]{\csname bibitem#1\endcsname}%
\let\auto@bib@innerbib\@empty
\bibitem [{\citenamefont {Vicsek}\ and\ \citenamefont
  {Zafeiris}(2012)}]{Vicsek2012}%
  \BibitemOpen
  \bibfield  {author} {\bibinfo {author} {\bibfnamefont {T.}~\bibnamefont
  {Vicsek}}\ and\ \bibinfo {author} {\bibfnamefont {A.}~\bibnamefont
  {Zafeiris}},\ }\href {\doibase 10.1016/j.physrep.2012.03.004} {\bibfield
  {journal} {\bibinfo  {journal} {Phys. Rep.}\ }\textbf {\bibinfo {volume}
  {517}},\ \bibinfo {pages} {71} (\bibinfo {year} {2012})}\BibitemShut
  {NoStop}%
\bibitem [{\citenamefont {Sumpter}(2010)}]{sumpter2010}%
  \BibitemOpen
  \bibfield  {author} {\bibinfo {author} {\bibfnamefont {D.~J.}\ \bibnamefont
  {Sumpter}},\ }\href@noop {} {\emph {\bibinfo {title} {Collective Animal
  Behavior}}}\ (\bibinfo  {publisher} {Princeton University Press},\ \bibinfo
  {address} {New Jersey},\ \bibinfo {year} {2010})\BibitemShut {NoStop}%
\bibitem [{\citenamefont {Ramaswamy}(2010)}]{Ramaswamy2010}%
  \BibitemOpen
  \bibfield  {author} {\bibinfo {author} {\bibfnamefont {S.}~\bibnamefont
  {Ramaswamy}},\ }\href {\doibase 10.1146/annurev-conmatphys-070909-104101}
  {\bibfield  {journal} {\bibinfo  {journal} {Annual Review of Condensed Matter
  Physics}\ }\textbf {\bibinfo {volume} {1}},\ \bibinfo {pages} {323} (\bibinfo
  {year} {2010})}\BibitemShut {NoStop}%
\bibitem [{\citenamefont {Rosenthal}\ \emph {et~al.}(2015)\citenamefont
  {Rosenthal}, \citenamefont {Twomey}, \citenamefont {Hartnett}, \citenamefont
  {Wu},\ and\ \citenamefont {Couzin}}]{Rosenthal2015}%
  \BibitemOpen
  \bibfield  {author} {\bibinfo {author} {\bibfnamefont {S.~B.}\ \bibnamefont
  {Rosenthal}}, \bibinfo {author} {\bibfnamefont {C.~R.}\ \bibnamefont
  {Twomey}}, \bibinfo {author} {\bibfnamefont {A.~T.}\ \bibnamefont
  {Hartnett}}, \bibinfo {author} {\bibfnamefont {H.~S.}\ \bibnamefont {Wu}}, \
  and\ \bibinfo {author} {\bibfnamefont {I.~D.}\ \bibnamefont {Couzin}},\
  }\href {\doibase 10.1073/pnas.1420068112} {\bibfield  {journal} {\bibinfo
  {journal} {Proc. Natl. Acad. Sci.}\ }\textbf {\bibinfo {volume} {112}},\
  \bibinfo {pages} {4690} (\bibinfo {year} {2015})}\BibitemShut {NoStop}%
\bibitem [{\citenamefont {Chen}\ \emph {et~al.}(2016)\citenamefont {Chen},
  \citenamefont {Vicsek}, \citenamefont {Liu}, \citenamefont {Zhou},\ and\
  \citenamefont {Zhang}}]{Chen2016}%
  \BibitemOpen
  \bibfield  {author} {\bibinfo {author} {\bibfnamefont {D.}~\bibnamefont
  {Chen}}, \bibinfo {author} {\bibfnamefont {T.}~\bibnamefont {Vicsek}},
  \bibinfo {author} {\bibfnamefont {X.}~\bibnamefont {Liu}}, \bibinfo {author}
  {\bibfnamefont {T.}~\bibnamefont {Zhou}}, \ and\ \bibinfo {author}
  {\bibfnamefont {H.}~\bibnamefont {Zhang}},\ }\href@noop {} {\bibfield
  {journal} {\bibinfo  {journal} {Europhys. Lett.}\ }\textbf {\bibinfo {volume}
  {114}},\ \bibinfo {pages} {60008} (\bibinfo {year} {2016})}\BibitemShut
  {NoStop}%
\bibitem [{\citenamefont {Calovi}\ \emph {et~al.}(2018)\citenamefont {Calovi},
  \citenamefont {Litchinko}, \citenamefont {Lecheval}, \citenamefont {Lopez},
  \citenamefont {P{\'e}rez~Escudero}, \citenamefont {Chat{\'e}}, \citenamefont
  {Sire},\ and\ \citenamefont {Theraulaz}}]{Calovi2018}%
  \BibitemOpen
  \bibfield  {author} {\bibinfo {author} {\bibfnamefont {D.~S.}\ \bibnamefont
  {Calovi}}, \bibinfo {author} {\bibfnamefont {A.}~\bibnamefont {Litchinko}},
  \bibinfo {author} {\bibfnamefont {V.}~\bibnamefont {Lecheval}}, \bibinfo
  {author} {\bibfnamefont {U.}~\bibnamefont {Lopez}}, \bibinfo {author}
  {\bibfnamefont {A.}~\bibnamefont {P{\'e}rez~Escudero}}, \bibinfo {author}
  {\bibfnamefont {H.}~\bibnamefont {Chat{\'e}}}, \bibinfo {author}
  {\bibfnamefont {C.}~\bibnamefont {Sire}}, \ and\ \bibinfo {author}
  {\bibfnamefont {G.}~\bibnamefont {Theraulaz}},\ }\href {\doibase
  10.1371/journal.pcbi.1005933} {\bibfield  {journal} {\bibinfo  {journal}
  {PLOS Computational Biology}\ }\textbf {\bibinfo {volume} {14}},\ \bibinfo
  {pages} {1} (\bibinfo {year} {2018})}\BibitemShut {NoStop}%
\bibitem [{\citenamefont {Brambilla}\ \emph {et~al.}(2013)\citenamefont
  {Brambilla}, \citenamefont {Ferrante}, \citenamefont {Birattari},\ and\
  \citenamefont {Dorigo}}]{Brambilla2013}%
  \BibitemOpen
  \bibfield  {author} {\bibinfo {author} {\bibfnamefont {M.}~\bibnamefont
  {Brambilla}}, \bibinfo {author} {\bibfnamefont {E.}~\bibnamefont {Ferrante}},
  \bibinfo {author} {\bibfnamefont {M.}~\bibnamefont {Birattari}}, \ and\
  \bibinfo {author} {\bibfnamefont {M.}~\bibnamefont {Dorigo}},\ }\href
  {\doibase 10.1007/s11721-012-0075-2} {\bibfield  {journal} {\bibinfo
  {journal} {Swarm Intelligence}\ }\textbf {\bibinfo {volume} {7}},\ \bibinfo
  {pages} {1} (\bibinfo {year} {2013})}\BibitemShut {NoStop}%
\bibitem [{\citenamefont {Krause}\ and\ \citenamefont
  {Ruxton}(2002)}]{Krause2002}%
  \BibitemOpen
  \bibfield  {author} {\bibinfo {author} {\bibfnamefont {J.}~\bibnamefont
  {Krause}}\ and\ \bibinfo {author} {\bibfnamefont {G.}~\bibnamefont
  {Ruxton}},\ }\href@noop {} {\emph {\bibinfo {title} {Living in Groups}}}\
  (\bibinfo  {publisher} {Oxford University Press},\ \bibinfo {address}
  {Oxford},\ \bibinfo {year} {2002})\BibitemShut {NoStop}%
\bibitem [{\citenamefont {Procaccini}\ \emph {et~al.}(2011)\citenamefont
  {Procaccini}, \citenamefont {Orlandi}, \citenamefont {Cavagna}, \citenamefont
  {Giardina}, \citenamefont {Zoratto}, \citenamefont {Santucci}, \citenamefont
  {Chiarotti}, \citenamefont {Hemelrijk}, \citenamefont {Alleva}, \citenamefont
  {Parisi},\ and\ \citenamefont {Carere}}]{Procaccini2011}%
  \BibitemOpen
  \bibfield  {author} {\bibinfo {author} {\bibfnamefont {A.}~\bibnamefont
  {Procaccini}}, \bibinfo {author} {\bibfnamefont {A.}~\bibnamefont {Orlandi}},
  \bibinfo {author} {\bibfnamefont {A.}~\bibnamefont {Cavagna}}, \bibinfo
  {author} {\bibfnamefont {I.}~\bibnamefont {Giardina}}, \bibinfo {author}
  {\bibfnamefont {F.}~\bibnamefont {Zoratto}}, \bibinfo {author} {\bibfnamefont
  {D.}~\bibnamefont {Santucci}}, \bibinfo {author} {\bibfnamefont
  {F.}~\bibnamefont {Chiarotti}}, \bibinfo {author} {\bibfnamefont {C.~K.}\
  \bibnamefont {Hemelrijk}}, \bibinfo {author} {\bibfnamefont {E.}~\bibnamefont
  {Alleva}}, \bibinfo {author} {\bibfnamefont {G.}~\bibnamefont {Parisi}}, \
  and\ \bibinfo {author} {\bibfnamefont {C.}~\bibnamefont {Carere}},\ }\href
  {\doibase 10.1016/j.anbehav.2011.07.006} {\bibfield  {journal} {\bibinfo
  {journal} {Animal Behaviour}\ }\textbf {\bibinfo {volume} {82}},\ \bibinfo
  {pages} {759} (\bibinfo {year} {2011})}\BibitemShut {NoStop}%
\bibitem [{\citenamefont {Ginelli}\ \emph {et~al.}(2015)\citenamefont
  {Ginelli}, \citenamefont {Peruani}, \citenamefont {Pillot}, \citenamefont
  {Chat{\'{e}}}, \citenamefont {Theraulaz},\ and\ \citenamefont
  {Bon}}]{Ginelli2015}%
  \BibitemOpen
  \bibfield  {author} {\bibinfo {author} {\bibfnamefont {F.}~\bibnamefont
  {Ginelli}}, \bibinfo {author} {\bibfnamefont {F.}~\bibnamefont {Peruani}},
  \bibinfo {author} {\bibfnamefont {M.-H.}\ \bibnamefont {Pillot}}, \bibinfo
  {author} {\bibfnamefont {H.}~\bibnamefont {Chat{\'{e}}}}, \bibinfo {author}
  {\bibfnamefont {G.}~\bibnamefont {Theraulaz}}, \ and\ \bibinfo {author}
  {\bibfnamefont {R.}~\bibnamefont {Bon}},\ }\href {\doibase
  10.1073/pnas.1503749112} {\bibfield  {journal} {\bibinfo  {journal}
  {Proceedings of the National Academy of Sciences}\ }\textbf {\bibinfo
  {volume} {112}},\ \bibinfo {pages} {12729} (\bibinfo {year}
  {2015})}\BibitemShut {NoStop}%
\bibitem [{\citenamefont {Couzin}\ \emph {et~al.}(2005)\citenamefont {Couzin},
  \citenamefont {Krause}, \citenamefont {Franks},\ and\ \citenamefont
  {Levin}}]{Couzin2005}%
  \BibitemOpen
  \bibfield  {author} {\bibinfo {author} {\bibfnamefont {I.~D.}\ \bibnamefont
  {Couzin}}, \bibinfo {author} {\bibfnamefont {J.}~\bibnamefont {Krause}},
  \bibinfo {author} {\bibfnamefont {N.~R.}\ \bibnamefont {Franks}}, \ and\
  \bibinfo {author} {\bibfnamefont {S.~A.}\ \bibnamefont {Levin}},\ }\href
  {\doibase 10.1038/nature03236} {\bibfield  {journal} {\bibinfo  {journal}
  {Nature}\ }\textbf {\bibinfo {volume} {433}},\ \bibinfo {pages} {513}
  (\bibinfo {year} {2005})}\BibitemShut {NoStop}%
\bibitem [{\citenamefont {Aub{\'{e}}}\ and\ \citenamefont
  {Shield}(2004)}]{Aube2004}%
  \BibitemOpen
  \bibfield  {author} {\bibinfo {author} {\bibfnamefont {F.}~\bibnamefont
  {Aub{\'{e}}}}\ and\ \bibinfo {author} {\bibfnamefont {R.}~\bibnamefont
  {Shield}},\ }in\ \href {\doibase 10.1007/978-3-540-30479-1_62} {\emph
  {\bibinfo {booktitle} {Lect. Notes Comput. Sci.}}},\ Vol.\ \bibinfo {volume}
  {3305}\ (\bibinfo  {publisher} {Springer Berlin / Heidelberg},\ \bibinfo
  {year} {2004})\ pp.\ \bibinfo {pages} {601--611}\BibitemShut {NoStop}%
\bibitem [{\citenamefont {Shen}(2008)}]{doi:10.1137/060673254}%
  \BibitemOpen
  \bibfield  {author} {\bibinfo {author} {\bibfnamefont {J.~J.}\ \bibnamefont
  {Shen}},\ }\href {\doibase 10.1137/060673254} {\bibfield  {journal} {\bibinfo
   {journal} {SIAM Journal on Applied Mathematics}\ }\textbf {\bibinfo {volume}
  {68}},\ \bibinfo {pages} {694} (\bibinfo {year} {2008})},\ \Eprint
  {http://arxiv.org/abs/https://doi.org/10.1137/060673254}
  {https://doi.org/10.1137/060673254} \BibitemShut {NoStop}%
\bibitem [{\citenamefont {Pearce}\ and\ \citenamefont
  {Giomi}(2016)}]{Pearce2016}%
  \BibitemOpen
  \bibfield  {author} {\bibinfo {author} {\bibfnamefont {D.~J.~G.}\
  \bibnamefont {Pearce}}\ and\ \bibinfo {author} {\bibfnamefont
  {L.}~\bibnamefont {Giomi}},\ }\href {\doibase 10.1103/PhysRevE.94.022612}
  {\bibfield  {journal} {\bibinfo  {journal} {Phys. Rev. E}\ }\textbf {\bibinfo
  {volume} {94}},\ \bibinfo {pages} {022612} (\bibinfo {year}
  {2016})}\BibitemShut {NoStop}%
\bibitem [{\citenamefont {Kyriakopoulos}\ \emph {et~al.}(2016)\citenamefont
  {Kyriakopoulos}, \citenamefont {Ginelli},\ and\ \citenamefont
  {Toner}}]{Kyriakopoulos_2016}%
  \BibitemOpen
  \bibfield  {author} {\bibinfo {author} {\bibfnamefont {N.}~\bibnamefont
  {Kyriakopoulos}}, \bibinfo {author} {\bibfnamefont {F.}~\bibnamefont
  {Ginelli}}, \ and\ \bibinfo {author} {\bibfnamefont {J.}~\bibnamefont
  {Toner}},\ }\href {\doibase 10.1088/1367-2630/18/7/073039} {\bibfield
  {journal} {\bibinfo  {journal} {New Journal of Physics}\ }\textbf {\bibinfo
  {volume} {18}},\ \bibinfo {pages} {73039} (\bibinfo {year}
  {2016})}\BibitemShut {NoStop}%
\bibitem [{\citenamefont {Smith}\ \emph {et~al.}(2016)\citenamefont {Smith},
  \citenamefont {Gavrilets}, \citenamefont {Mulder}, \citenamefont {Hooper},
  \citenamefont {Mouden}, \citenamefont {Nettle}, \citenamefont {Hauert},
  \citenamefont {Hill}, \citenamefont {Perry}, \citenamefont {Pusey},
  \citenamefont {van Vugt},\ and\ \citenamefont {Smith}}]{Smith2016}%
  \BibitemOpen
  \bibfield  {author} {\bibinfo {author} {\bibfnamefont {J.~E.}\ \bibnamefont
  {Smith}}, \bibinfo {author} {\bibfnamefont {S.}~\bibnamefont {Gavrilets}},
  \bibinfo {author} {\bibfnamefont {M.~B.}\ \bibnamefont {Mulder}}, \bibinfo
  {author} {\bibfnamefont {P.~L.}\ \bibnamefont {Hooper}}, \bibinfo {author}
  {\bibfnamefont {C.~E.}\ \bibnamefont {Mouden}}, \bibinfo {author}
  {\bibfnamefont {D.}~\bibnamefont {Nettle}}, \bibinfo {author} {\bibfnamefont
  {C.}~\bibnamefont {Hauert}}, \bibinfo {author} {\bibfnamefont
  {K.}~\bibnamefont {Hill}}, \bibinfo {author} {\bibfnamefont {S.}~\bibnamefont
  {Perry}}, \bibinfo {author} {\bibfnamefont {A.~E.}\ \bibnamefont {Pusey}},
  \bibinfo {author} {\bibfnamefont {M.}~\bibnamefont {van Vugt}}, \ and\
  \bibinfo {author} {\bibfnamefont {E.~A.}\ \bibnamefont {Smith}},\ }\href
  {\doibase 10.1016/j.tree.2015.09.013} {\bibfield  {journal} {\bibinfo
  {journal} {Trends in Ecology {\&} Evolution}\ }\textbf {\bibinfo {volume}
  {31}},\ \bibinfo {pages} {54} (\bibinfo {year} {2016})}\BibitemShut {NoStop}%
\bibitem [{\citenamefont {Nagy}\ \emph {et~al.}(2010)\citenamefont {Nagy},
  \citenamefont {Akos}, \citenamefont {Biro}, \citenamefont {Vicsek},
  \citenamefont {{\'{A}}kos}, \citenamefont {Biro},\ and\ \citenamefont
  {Vicsek}}]{Nagy2010}%
  \BibitemOpen
  \bibfield  {author} {\bibinfo {author} {\bibfnamefont {M.}~\bibnamefont
  {Nagy}}, \bibinfo {author} {\bibfnamefont {Z.}~\bibnamefont {Akos}}, \bibinfo
  {author} {\bibfnamefont {D.}~\bibnamefont {Biro}}, \bibinfo {author}
  {\bibfnamefont {T.}~\bibnamefont {Vicsek}}, \bibinfo {author} {\bibfnamefont
  {Z.}~\bibnamefont {{\'{A}}kos}}, \bibinfo {author} {\bibfnamefont
  {D.}~\bibnamefont {Biro}}, \ and\ \bibinfo {author} {\bibfnamefont
  {T.}~\bibnamefont {Vicsek}},\ }\href {\doibase 10.1038/nature08891}
  {\bibfield  {journal} {\bibinfo  {journal} {Nature}\ }\textbf {\bibinfo
  {volume} {464}},\ \bibinfo {pages} {890} (\bibinfo {year} {2010})},\ \Eprint
  {http://arxiv.org/abs/1010.5394} {arXiv:1010.5394} \BibitemShut {NoStop}%
\bibitem [{\citenamefont {Flack}\ \emph {et~al.}(2012)\citenamefont {Flack},
  \citenamefont {Pettit}, \citenamefont {Freeman}, \citenamefont {Guilford},\
  and\ \citenamefont {Biro}}]{Flack2012}%
  \BibitemOpen
  \bibfield  {author} {\bibinfo {author} {\bibfnamefont {A.}~\bibnamefont
  {Flack}}, \bibinfo {author} {\bibfnamefont {B.}~\bibnamefont {Pettit}},
  \bibinfo {author} {\bibfnamefont {R.}~\bibnamefont {Freeman}}, \bibinfo
  {author} {\bibfnamefont {T.}~\bibnamefont {Guilford}}, \ and\ \bibinfo
  {author} {\bibfnamefont {D.}~\bibnamefont {Biro}},\ }\href {\doibase
  10.1016/j.anbehav.2011.12.018} {\bibfield  {journal} {\bibinfo  {journal}
  {Animal Behaviour}\ }\textbf {\bibinfo {volume} {83}},\ \bibinfo {pages}
  {703} (\bibinfo {year} {2012})}\BibitemShut {NoStop}%
\bibitem [{\citenamefont {Nagy}\ \emph {et~al.}(2013)\citenamefont {Nagy},
  \citenamefont {V{\'{a}}s{\'{a}}rhelyi}, \citenamefont {Pettit}, \citenamefont
  {Roberts-Mariani}, \citenamefont {Vicsek},\ and\ \citenamefont
  {Biro}}]{Nagy2013}%
  \BibitemOpen
  \bibfield  {author} {\bibinfo {author} {\bibfnamefont {M.}~\bibnamefont
  {Nagy}}, \bibinfo {author} {\bibfnamefont {G.}~\bibnamefont
  {V{\'{a}}s{\'{a}}rhelyi}}, \bibinfo {author} {\bibfnamefont {B.}~\bibnamefont
  {Pettit}}, \bibinfo {author} {\bibfnamefont {I.}~\bibnamefont
  {Roberts-Mariani}}, \bibinfo {author} {\bibfnamefont {T.}~\bibnamefont
  {Vicsek}}, \ and\ \bibinfo {author} {\bibfnamefont {D.}~\bibnamefont
  {Biro}},\ }\href {\doibase 10.1073/pnas.1305552110} {\bibfield  {journal}
  {\bibinfo  {journal} {Proceedings of the National Academy of Sciences of the
  United States of America}\ }\textbf {\bibinfo {volume} {110}},\ \bibinfo
  {pages} {13049} (\bibinfo {year} {2013})}\BibitemShut {NoStop}%
\bibitem [{\citenamefont {Couzin}\ \emph {et~al.}(2011)\citenamefont {Couzin},
  \citenamefont {Ioannou}, \citenamefont {Demirel}, \citenamefont {Gross},
  \citenamefont {Torney}, \citenamefont {Hartnett}, \citenamefont {Conradt},
  \citenamefont {Levin},\ and\ \citenamefont {Leonard}}]{Couzin2011}%
  \BibitemOpen
  \bibfield  {author} {\bibinfo {author} {\bibfnamefont {I.~D.}\ \bibnamefont
  {Couzin}}, \bibinfo {author} {\bibfnamefont {C.~C.}\ \bibnamefont {Ioannou}},
  \bibinfo {author} {\bibfnamefont {G.}~\bibnamefont {Demirel}}, \bibinfo
  {author} {\bibfnamefont {T.}~\bibnamefont {Gross}}, \bibinfo {author}
  {\bibfnamefont {C.~J.}\ \bibnamefont {Torney}}, \bibinfo {author}
  {\bibfnamefont {a.}~\bibnamefont {Hartnett}}, \bibinfo {author}
  {\bibfnamefont {L.}~\bibnamefont {Conradt}}, \bibinfo {author} {\bibfnamefont
  {S.~a.}\ \bibnamefont {Levin}}, \ and\ \bibinfo {author} {\bibfnamefont
  {N.~E.}\ \bibnamefont {Leonard}},\ }\href {\doibase 10.1126/science.1210280}
  {\bibfield  {journal} {\bibinfo  {journal} {Science}\ }\textbf {\bibinfo
  {volume} {334}},\ \bibinfo {pages} {1578} (\bibinfo {year}
  {2011})}\BibitemShut {NoStop}%
\bibitem [{\citenamefont {Ward}\ \emph {et~al.}(2011)\citenamefont {Ward},
  \citenamefont {Herbert-Read}, \citenamefont {Sumpter},\ and\ \citenamefont
  {Krause}}]{Ward08022011}%
  \BibitemOpen
  \bibfield  {author} {\bibinfo {author} {\bibfnamefont {A.~J.~W.}\
  \bibnamefont {Ward}}, \bibinfo {author} {\bibfnamefont {J.~E.}\ \bibnamefont
  {Herbert-Read}}, \bibinfo {author} {\bibfnamefont {D.~J.~T.}\ \bibnamefont
  {Sumpter}}, \ and\ \bibinfo {author} {\bibfnamefont {J.}~\bibnamefont
  {Krause}},\ }\href {\doibase 10.1073/pnas.1007102108} {\bibfield  {journal}
  {\bibinfo  {journal} {Proceedings of the National Academy of Sciences}\
  }\textbf {\bibinfo {volume} {108}},\ \bibinfo {pages} {2312} (\bibinfo {year}
  {2011})}\BibitemShut {NoStop}%
\bibitem [{\citenamefont {Fisher}(1998)}]{Fisher1998}%
  \BibitemOpen
  \bibfield  {author} {\bibinfo {author} {\bibfnamefont {D.~S.}\ \bibnamefont
  {Fisher}},\ }\href {\doibase 10.1016/S0370-1573(98)00008-8} {\bibfield
  {journal} {\bibinfo  {journal} {Physics Reports}\ }\textbf {\bibinfo {volume}
  {301}},\ \bibinfo {pages} {113} (\bibinfo {year} {1998})},\ \Eprint
  {http://arxiv.org/abs/9711179} {arXiv:9711179 [cond-mat]} \BibitemShut
  {NoStop}%
\bibitem [{\citenamefont {Zapperi}\ \emph {et~al.}(1998)\citenamefont
  {Zapperi}, \citenamefont {Cizeau}, \citenamefont {Durin},\ and\ \citenamefont
  {Stanley}}]{PhysRevB.58.6353}%
  \BibitemOpen
  \bibfield  {author} {\bibinfo {author} {\bibfnamefont {S.}~\bibnamefont
  {Zapperi}}, \bibinfo {author} {\bibfnamefont {P.}~\bibnamefont {Cizeau}},
  \bibinfo {author} {\bibfnamefont {G.}~\bibnamefont {Durin}}, \ and\ \bibinfo
  {author} {\bibfnamefont {H.~E.}\ \bibnamefont {Stanley}},\ }\href {\doibase
  10.1103/PhysRevB.58.6353} {\bibfield  {journal} {\bibinfo  {journal} {Phys.
  Rev. B}\ }\textbf {\bibinfo {volume} {58}},\ \bibinfo {pages} {6353}
  (\bibinfo {year} {1998})}\BibitemShut {NoStop}%
\bibitem [{\citenamefont {Altshuler}\ and\ \citenamefont
  {Johansen}(2004)}]{RevModPhys.76.471}%
  \BibitemOpen
  \bibfield  {author} {\bibinfo {author} {\bibfnamefont {E.}~\bibnamefont
  {Altshuler}}\ and\ \bibinfo {author} {\bibfnamefont {T.~H.}\ \bibnamefont
  {Johansen}},\ }\href {\doibase 10.1103/RevModPhys.76.471} {\bibfield
  {journal} {\bibinfo  {journal} {Rev. Mod. Phys.}\ }\textbf {\bibinfo {volume}
  {76}},\ \bibinfo {pages} {471} (\bibinfo {year} {2004})}\BibitemShut
  {NoStop}%
\bibitem [{\citenamefont {Miguel}\ \emph {et~al.}(2001)\citenamefont {Miguel},
  \citenamefont {Vespignani}, \citenamefont {Zapperi}, \citenamefont {Weiss},\
  and\ \citenamefont {Grasso}}]{Miguel2001}%
  \BibitemOpen
  \bibfield  {author} {\bibinfo {author} {\bibfnamefont {M.~C.}\ \bibnamefont
  {Miguel}}, \bibinfo {author} {\bibfnamefont {A.}~\bibnamefont {Vespignani}},
  \bibinfo {author} {\bibfnamefont {S.}~\bibnamefont {Zapperi}}, \bibinfo
  {author} {\bibfnamefont {J.}~\bibnamefont {Weiss}}, \ and\ \bibinfo {author}
  {\bibfnamefont {J.-R.}\ \bibnamefont {Grasso}},\ }\href {\doibase
  10.1038/35070524} {\bibfield  {journal} {\bibinfo  {journal} {Nature}\
  }\textbf {\bibinfo {volume} {410}} (\bibinfo {year} {2001}),\
  10.1038/35070524}\BibitemShut {NoStop}%
\bibitem [{\citenamefont {Zapperi}\ \emph {et~al.}(1999)\citenamefont
  {Zapperi}, \citenamefont {Ray}, \citenamefont {Stanley},\ and\ \citenamefont
  {Vespignani}}]{PhysRevE.59.5049}%
  \BibitemOpen
  \bibfield  {author} {\bibinfo {author} {\bibfnamefont {S.}~\bibnamefont
  {Zapperi}}, \bibinfo {author} {\bibfnamefont {P.}~\bibnamefont {Ray}},
  \bibinfo {author} {\bibfnamefont {H.~E.}\ \bibnamefont {Stanley}}, \ and\
  \bibinfo {author} {\bibfnamefont {A.}~\bibnamefont {Vespignani}},\ }\href
  {\doibase 10.1103/PhysRevE.59.5049} {\bibfield  {journal} {\bibinfo
  {journal} {Phys. Rev. E}\ }\textbf {\bibinfo {volume} {59}},\ \bibinfo
  {pages} {5049} (\bibinfo {year} {1999})}\BibitemShut {NoStop}%
\bibitem [{\citenamefont {Kawamura}\ \emph {et~al.}(2012)\citenamefont
  {Kawamura}, \citenamefont {Hatano}, \citenamefont {Kato}, \citenamefont
  {Biswas},\ and\ \citenamefont {Chakrabarti}}]{RevModPhys.84.839}%
  \BibitemOpen
  \bibfield  {author} {\bibinfo {author} {\bibfnamefont {H.}~\bibnamefont
  {Kawamura}}, \bibinfo {author} {\bibfnamefont {T.}~\bibnamefont {Hatano}},
  \bibinfo {author} {\bibfnamefont {N.}~\bibnamefont {Kato}}, \bibinfo {author}
  {\bibfnamefont {S.}~\bibnamefont {Biswas}}, \ and\ \bibinfo {author}
  {\bibfnamefont {B.~K.}\ \bibnamefont {Chakrabarti}},\ }\href {\doibase
  10.1103/RevModPhys.84.839} {\bibfield  {journal} {\bibinfo  {journal} {Rev.
  Mod. Phys.}\ }\textbf {\bibinfo {volume} {84}},\ \bibinfo {pages} {839}
  (\bibinfo {year} {2012})}\BibitemShut {NoStop}%
\bibitem [{\citenamefont {Vicsek}\ \emph {et~al.}(1995)\citenamefont {Vicsek},
  \citenamefont {Czirok}, \citenamefont {Ben-Jacob}, \citenamefont {Cohen},\
  and\ \citenamefont {Shochet}}]{vicsek1995}%
  \BibitemOpen
  \bibfield  {author} {\bibinfo {author} {\bibfnamefont {T.}~\bibnamefont
  {Vicsek}}, \bibinfo {author} {\bibfnamefont {A.}~\bibnamefont {Czirok}},
  \bibinfo {author} {\bibfnamefont {E.}~\bibnamefont {Ben-Jacob}}, \bibinfo
  {author} {\bibfnamefont {I.}~\bibnamefont {Cohen}}, \ and\ \bibinfo {author}
  {\bibfnamefont {O.}~\bibnamefont {Shochet}},\ }\href {\doibase
  10.1103/PhysRevLett.75.1226} {\bibfield  {journal} {\bibinfo  {journal}
  {Phys. Rev. Lett.}\ }\textbf {\bibinfo {volume} {75}},\ \bibinfo {pages}
  {1226} (\bibinfo {year} {1995})}\BibitemShut {NoStop}%
\bibitem [{\citenamefont {Pruessner}(2012)}]{pruessner2012}%
  \BibitemOpen
  \bibfield  {author} {\bibinfo {author} {\bibfnamefont {G.}~\bibnamefont
  {Pruessner}},\ }\href {https://books.google.es/books?id=TXKcpGqMSDYC} {\emph
  {\bibinfo {title} {Self-Organised Criticality: Theory, Models and
  Characterisation}}},\ Self-organised Criticality: Theory, Models, and
  Characterisation\ (\bibinfo  {publisher} {Cambridge University Press},\
  \bibinfo {address} {Cambridge, UK},\ \bibinfo {year} {2012})\BibitemShut
  {NoStop}%
\bibitem [{\citenamefont {M\'endez}\ \emph {et~al.}(2014)\citenamefont
  {M\'endez}, \citenamefont {Campos},\ and\ \citenamefont
  {Bartumeus}}]{mendez14}%
  \BibitemOpen
  \bibfield  {author} {\bibinfo {author} {\bibfnamefont {V.}~\bibnamefont
  {M\'endez}}, \bibinfo {author} {\bibfnamefont {D.}~\bibnamefont {Campos}}, \
  and\ \bibinfo {author} {\bibfnamefont {F.}~\bibnamefont {Bartumeus}},\
  }\href@noop {} {\emph {\bibinfo {title} {Stochastic Foundations in Movement
  Ecology}}}\ (\bibinfo  {publisher} {Springer Verlag},\ \bibinfo {address}
  {Berlin, Heidelberg},\ \bibinfo {year} {2014})\BibitemShut {NoStop}%
\bibitem [{\citenamefont {Gimeno}\ \emph {et~al.}(2016)\citenamefont {Gimeno},
  \citenamefont {Quera}, \citenamefont {Beltran},\ and\ \citenamefont
  {Dolado}}]{gimeno2016}%
  \BibitemOpen
  \bibfield  {author} {\bibinfo {author} {\bibfnamefont {E.}~\bibnamefont
  {Gimeno}}, \bibinfo {author} {\bibfnamefont {V.}~\bibnamefont {Quera}},
  \bibinfo {author} {\bibfnamefont {F.~S.}\ \bibnamefont {Beltran}}, \ and\
  \bibinfo {author} {\bibfnamefont {R.}~\bibnamefont {Dolado}},\ }\href@noop {}
  {\bibfield  {journal} {\bibinfo  {journal} {Journal of Comparative
  Psychology}\ }\textbf {\bibinfo {volume} {130}},\ \bibinfo {pages} {358}
  (\bibinfo {year} {2016})}\BibitemShut {NoStop}%
\bibitem [{\citenamefont {Press}\ and\ \citenamefont
  {Teukolsky}(1990)}]{doi:10.1063/1.4822961}%
  \BibitemOpen
  \bibfield  {author} {\bibinfo {author} {\bibfnamefont {W.~H.}\ \bibnamefont
  {Press}}\ and\ \bibinfo {author} {\bibfnamefont {S.~A.}\ \bibnamefont
  {Teukolsky}},\ }\href {\doibase 10.1063/1.4822961} {\bibfield  {journal}
  {\bibinfo  {journal} {Computers in Physics}\ }\textbf {\bibinfo {volume}
  {4}},\ \bibinfo {pages} {669} (\bibinfo {year} {1990})}\BibitemShut {NoStop}%
\bibitem [{\citenamefont {Fornberg}(1988)}]{Fornberg1988}%
  \BibitemOpen
  \bibfield  {author} {\bibinfo {author} {\bibfnamefont {B.}~\bibnamefont
  {Fornberg}},\ }\href {\doibase 10.1090/S0025-5718-1988-0935077-0} {\bibfield
  {journal} {\bibinfo  {journal} {Mathematics of Computation}\ }\textbf
  {\bibinfo {volume} {51}},\ \bibinfo {pages} {699} (\bibinfo {year}
  {1988})}\BibitemShut {NoStop}%
\bibitem [{\citenamefont {Laurson}\ and\ \citenamefont
  {Alava}(2006)}]{Laurson2006}%
  \BibitemOpen
  \bibfield  {author} {\bibinfo {author} {\bibfnamefont {L.}~\bibnamefont
  {Laurson}}\ and\ \bibinfo {author} {\bibfnamefont {M.~J.}\ \bibnamefont
  {Alava}},\ }\href@noop {} {\bibfield  {journal} {\bibinfo  {journal} {Phys.
  Rev. E}\ }\textbf {\bibinfo {volume} {74}} (\bibinfo {year}
  {2006})}\BibitemShut {NoStop}%
\bibitem [{\citenamefont {Laurson}\ \emph {et~al.}(2009)\citenamefont
  {Laurson}, \citenamefont {Illa},\ and\ \citenamefont {Alava}}]{Laurson2009}%
  \BibitemOpen
  \bibfield  {author} {\bibinfo {author} {\bibfnamefont {L.}~\bibnamefont
  {Laurson}}, \bibinfo {author} {\bibfnamefont {X.}~\bibnamefont {Illa}}, \
  and\ \bibinfo {author} {\bibfnamefont {M.~J.}\ \bibnamefont {Alava}},\
  }\href@noop {} {\bibfield  {journal} {\bibinfo  {journal} {J. Stat. Mech.:
  Theo. Exp.}\ }\textbf {\bibinfo {volume} {P01019}} (\bibinfo {year}
  {2009})}\BibitemShut {NoStop}%
\bibitem [{\citenamefont {Yeomans}(1992)}]{yeomans}%
  \BibitemOpen
  \bibfield  {author} {\bibinfo {author} {\bibfnamefont {J.~M.}\ \bibnamefont
  {Yeomans}},\ }\href@noop {} {\emph {\bibinfo {title} {Statistical mechanics
  of phase transitions}}}\ (\bibinfo  {publisher} {Oxford University Press},\
  \bibinfo {address} {Oxford},\ \bibinfo {year} {1992})\BibitemShut {NoStop}%
\bibitem [{\citenamefont {Cardy}(1996)}]{cardy_1996}%
  \BibitemOpen
  \bibfield  {author} {\bibinfo {author} {\bibfnamefont {J.}~\bibnamefont
  {Cardy}},\ }\href {\doibase 10.1017/CBO9781316036440} {\emph {\bibinfo
  {title} {Scaling and Renormalization in Statistical Physics}}},\ Cambridge
  Lecture Notes in Physics\ (\bibinfo  {publisher} {Cambridge University
  Press},\ \bibinfo {year} {1996})\BibitemShut {NoStop}%
\bibitem [{\citenamefont {Strandburg-Peshkin}\ \emph
  {et~al.}(2018)\citenamefont {Strandburg-Peshkin}, \citenamefont
  {Papageorgiou}, \citenamefont {Crofoot},\ and\ \citenamefont
  {Farine}}]{Strandburg-Peshkin.2018}%
  \BibitemOpen
  \bibfield  {author} {\bibinfo {author} {\bibfnamefont {A.}~\bibnamefont
  {Strandburg-Peshkin}}, \bibinfo {author} {\bibfnamefont {D.}~\bibnamefont
  {Papageorgiou}}, \bibinfo {author} {\bibfnamefont {M.~C.}\ \bibnamefont
  {Crofoot}}, \ and\ \bibinfo {author} {\bibfnamefont {D.~R.}\ \bibnamefont
  {Farine}},\ }\href {\doibase 10.1098/rstb.2017.0006} {\bibfield  {journal}
  {\bibinfo  {journal} {Philosophical Transactions of the Royal Society B:
  Biological Sciences}\ }\textbf {\bibinfo {volume} {373}},\ \bibinfo {pages}
  {20170006} (\bibinfo {year} {2018})}\BibitemShut {NoStop}%
\bibitem [{\citenamefont {Ginelli}(2016)}]{Ginelli2016}%
  \BibitemOpen
  \bibfield  {author} {\bibinfo {author} {\bibfnamefont {F.}~\bibnamefont
  {Ginelli}},\ }\href {\doibase 10.1140/epjst/e2016-60066-8} {\bibfield
  {journal} {\bibinfo  {journal} {Eur. Phys. J. Spec. Top.}\ }\textbf {\bibinfo
  {volume} {225}},\ \bibinfo {pages} {2099} (\bibinfo {year}
  {2016})}\BibitemShut {NoStop}%
\bibitem [{\citenamefont {Zumaya}\ \emph {et~al.}(2018)\citenamefont {Zumaya},
  \citenamefont {Larralde},\ and\ \citenamefont {Aldana}}]{Zumaya.2018}%
  \BibitemOpen
  \bibfield  {author} {\bibinfo {author} {\bibfnamefont {M.}~\bibnamefont
  {Zumaya}}, \bibinfo {author} {\bibfnamefont {H.}~\bibnamefont {Larralde}}, \
  and\ \bibinfo {author} {\bibfnamefont {M.}~\bibnamefont {Aldana}},\ }\href
  {\doibase 10.1038/s41598-018-34208-x} {\bibfield  {journal} {\bibinfo
  {journal} {Scientific Reports}\ }\textbf {\bibinfo {volume} {8}},\ \bibinfo
  {pages} {15872} (\bibinfo {year} {2018})}\BibitemShut {NoStop}%
\bibitem [{\citenamefont {Cardy}(1988)}]{cardy88}%
  \BibitemOpen
  \bibinfo {editor} {\bibfnamefont {J.~L.}\ \bibnamefont {Cardy}},\ ed.,\
  \href@noop {} {\emph {\bibinfo {title} {Finite Size Scaling}}},\ \bibinfo
  {series} {Current Physics-Sources and Comments}, Vol.~\bibinfo {volume} {2}\
  (\bibinfo  {publisher} {North Holland},\ \bibinfo {address} {Amsterdam},\
  \bibinfo {year} {1988})\BibitemShut {NoStop}%
\bibitem [{\citenamefont {De~Menech}\ \emph {et~al.}(1998)\citenamefont
  {De~Menech}, \citenamefont {Stella},\ and\ \citenamefont
  {Tebaldi}}]{PhysRevE.58.R2677}%
  \BibitemOpen
  \bibfield  {author} {\bibinfo {author} {\bibfnamefont {M.}~\bibnamefont
  {De~Menech}}, \bibinfo {author} {\bibfnamefont {A.~L.}\ \bibnamefont
  {Stella}}, \ and\ \bibinfo {author} {\bibfnamefont {C.}~\bibnamefont
  {Tebaldi}},\ }\href {\doibase 10.1103/PhysRevE.58.R2677} {\bibfield
  {journal} {\bibinfo  {journal} {Phys. Rev. E}\ }\textbf {\bibinfo {volume}
  {58}},\ \bibinfo {pages} {R2677} (\bibinfo {year} {1998})}\BibitemShut
  {NoStop}%
\bibitem [{\citenamefont {Herbert-Read}\ \emph {et~al.}(2011)\citenamefont
  {Herbert-Read}, \citenamefont {Perna}, \citenamefont {Mann}, \citenamefont
  {Schaerf}, \citenamefont {Sumpter},\ and\ \citenamefont
  {Ward}}]{Herbert-Read.2011}%
  \BibitemOpen
  \bibfield  {author} {\bibinfo {author} {\bibfnamefont {J.~E.}\ \bibnamefont
  {Herbert-Read}}, \bibinfo {author} {\bibfnamefont {A.}~\bibnamefont {Perna}},
  \bibinfo {author} {\bibfnamefont {R.~P.}\ \bibnamefont {Mann}}, \bibinfo
  {author} {\bibfnamefont {T.~M.}\ \bibnamefont {Schaerf}}, \bibinfo {author}
  {\bibfnamefont {D.~J.~T.}\ \bibnamefont {Sumpter}}, \ and\ \bibinfo {author}
  {\bibfnamefont {A.~J.~W.}\ \bibnamefont {Ward}},\ }\href {\doibase
  10.1073/pnas.1109355108} {\bibfield  {journal} {\bibinfo  {journal}
  {Proceedings of the National Academy of Sciences}\ }\textbf {\bibinfo
  {volume} {108}},\ \bibinfo {pages} {18726} (\bibinfo {year}
  {2011})}\BibitemShut {NoStop}%
\bibitem [{\citenamefont {Katz}\ \emph {et~al.}(2011)\citenamefont {Katz},
  \citenamefont {Tunstr{\o}m}, \citenamefont {Ioannou}, \citenamefont {Huepe},\
  and\ \citenamefont {Couzin}}]{katz2011}%
  \BibitemOpen
  \bibfield  {author} {\bibinfo {author} {\bibfnamefont {Y.}~\bibnamefont
  {Katz}}, \bibinfo {author} {\bibfnamefont {K.}~\bibnamefont {Tunstr{\o}m}},
  \bibinfo {author} {\bibfnamefont {C.~C.}\ \bibnamefont {Ioannou}}, \bibinfo
  {author} {\bibfnamefont {C.}~\bibnamefont {Huepe}}, \ and\ \bibinfo {author}
  {\bibfnamefont {I.~D.}\ \bibnamefont {Couzin}},\ }\href {\doibase
  10.1073/pnas.1107583108} {\bibfield  {journal} {\bibinfo  {journal}
  {Proceedings of the National Academy of Sciences}\ }\textbf {\bibinfo
  {volume} {108}},\ \bibinfo {pages} {18720} (\bibinfo {year} {2011})},\
  \Eprint
  {http://arxiv.org/abs/https://www.pnas.org/content/108/46/18720.full.pdf}
  {https://www.pnas.org/content/108/46/18720.full.pdf} \BibitemShut {NoStop}%
\bibitem [{\citenamefont {Hoeffding}(1956)}]{10.1214/aoms/1177728178}%
  \BibitemOpen
  \bibfield  {author} {\bibinfo {author} {\bibfnamefont {W.}~\bibnamefont
  {Hoeffding}},\ }\href {\doibase 10.1214/aoms/1177728178} {\bibfield
  {journal} {\bibinfo  {journal} {The Annals of Mathematical Statistics}\
  }\textbf {\bibinfo {volume} {27}},\ \bibinfo {pages} {713 } (\bibinfo {year}
  {1956})}\BibitemShut {NoStop}%
\end{thebibliography}
\end{document}